\documentclass[a4paper,11pt]{article}
\pdfoutput=1 

\usepackage{jheppub} 

\usepackage[T1]{fontenc} 

\usepackage{amsmath}
\usepackage{amssymb}
\usepackage{epsfig,ulem,color,slashed}
\usepackage{multirow}
\allowdisplaybreaks  

\usepackage{todonotes}

\def \beq{\begin{equation}}
\def \eeq{\end{equation}}
\def \bea{\begin{eqnarray}}
\def \eea{\end{eqnarray}}

\newcommand{\m}[0]{\overline{m}}

\def \Op{Q}
\def \bm#1{\mbox{\boldmath$#1$\unboldmath}} 
\def \sp025{\hspace{0.25mm}}

\title{On new physics in $\bm{\Delta \Gamma_d}$}

\author[a]{Christoph Bobeth,}
\author[b,c]{Ulrich Haisch,}
\author[d]{Alexander Lenz,}
\author[d]{\\ Ben Pecjak}
\author[d]{and Gilberto Tetlalmatzi-Xolocotzi} 	

\affiliation[a]{Technische Universit{\"a}t M{\"u}nchen, Institute for Advanced Study, \\
85747 Garching, Germany}
\affiliation[b]{Rudolf Peierls Centre for Theoretical Physics, University of Oxford, \\
OX1 3PN Oxford, United Kingdom}
\affiliation[c]{CERN, Theory Division, \\ CH-1211 Geneva 23, Switzerland}
\affiliation[d]{Institute for Particle Physics Phenomenology, Durham University, \\ 
DH1 3LE Durham, United Kingdom}

\emailAdd{christoph.bobeth@ph.tum.de}
\emailAdd{u.haisch1@physics.ox.ac.uk}
\emailAdd{alexander.lenz@durham.ac.uk}
\emailAdd{ben.pecjak@durham.ac.uk}
\emailAdd{gilberto.tetlalmatzi-xolocotz@durham.ac.uk}

\abstract{ Motivated by the recent measurement of the dimuon asymmetry by the
  D{\O} collaboration, which could be interpreted as an enhanced decay rate
  difference in the neutral $B_d$-meson system, we investigate the possible size
  of new-physics contributions to $\Delta \Gamma_d$. In particular, we perform
  model-independent studies of non-standard effects associated to the
  dimension-six current-current operators $(\bar{d} p)(\bar p^{\hspace{0.25mm}
    \prime} b)$ with $p,p^\prime= u,c$ as well as $(\bar{d}b) (\bar\tau\tau)$.
  In both cases we find that for certain flavour or Lorentz structures of the
  operators sizable deviations of $\Delta \Gamma_d$ away from the Standard
  Model expectation cannot be excluded in a model-independent fashion.  }

\preprint{  FLAVOUR(267104)-ERC-66

  \vspace{-0.45cm}
 
  \begin{flushright}
  IPPP/14/29 \\
  DCPT/14/58
  \end{flushright}
}

\keywords{
  Mostly Weak Interactions: 
  B-Physics, CP violation, Rare Decays, Beyond Standard Model
}

\date{\today}

\begin{document} 

\maketitle

%
%

\section{Introduction}
\label{sec:intro}

The experimental measurements of the like-sign dimuon asymmetry by the D{\O}
collaboration in 2010 and 2011 \cite{Abazov:2010hv, Abazov:2010hj,
  Abazov:2011yk} triggered much interest in the flavour-physics community. If
the value of the measured asymmetry $A_{\rm CP}$ is interpreted solely as a
CP-violating effect in mixing of neutral $B_{d,s}$ mesons ($a_{sl}^{d,s}$),
\begin{equation}
  \label{Masterold}
  A_{\rm CP} \propto A_{sl}^b 
  \equiv C_d \hspace{0.5mm} a_{sl}^d + C_s \hspace{0.5mm} a_{sl}^s \,, 
\end{equation}
then using $C_d \simeq 0.59$ and $C_s\simeq 0.41$ the experimental number from
2011 \cite{Abazov:2011yk} deviates by $3.9\sp025\sigma$ from the latest Standard
Model~(SM) prediction \cite{Lenz:2011ti}.  One finds, however, that new-physics
contributions in $\Delta B = 2$ transitions alone cannot explain the large
central value of the like-sign dimuon asymmetry, since such a large enhancement would
violate model-independent bounds (see
e.g.~\cite{Dobrescu:2010rh,Lenz:2012mb}).

In \cite{Borissov:2013wwa} the interpretation of the like-sign dimuon asymmetry within the
SM was revisited. It was found that (\ref{Masterold}) should be modified to take
into account previously neglected contributions proportional to the decay rate
differences $\Delta \Gamma_d$ and $\Delta \Gamma_s$.  These arise from
interference of $B$-meson decays with and without mixing. 
The modified result reads
\begin{equation} 
  \label{Master}
  A_{\rm CP} \propto A_{sl}^b 
  + C_{\Gamma_d} \, \frac{\Delta \Gamma_d}{\Gamma_d} 
  + C_{\Gamma_s} \, \frac{\Delta \Gamma_s}{\Gamma_s} \,. 
\end{equation}
Numerical values for the coefficients $C_{\Gamma_d}$ and $C_{\Gamma_s}$ 
can be extracted from \cite{Borissov:2013wwa} and \cite{Abazov:2013uma} . 
The sign of  $C_{\Gamma_d}$
is such that a positive value of $\Delta \Gamma_d$ gives a negative contribution
to $A_{\rm CP}$ and $C_{\Gamma_s}$ turns out to be negligible.
It follows that the measured like-sign dimuon asymmetry is not simply proportional to the
semi-leptonic asymmetries $a_{sl}^{d,s}$, as assumed in \cite{Abazov:2010hv,
  Abazov:2010hj, Abazov:2011yk}.

Very recently the D\O{} collaboration presented a new measurement
\cite{Abazov:2013uma} of the coefficients $C_d$, $C_s$ and $C_{\Gamma_d}$ ($C_{\Gamma_s}$ was
neglected) and more importantly of the inclusive single-muon charge asymmetry
$a_{\rm CP}$ and the asymmetry $A_{\rm CP}$. The result is
\begin{align}
  \label{New} 
  a_{\rm CP} & = (-0.032 \pm 0.042 \pm 0.061) \% \,, &
  A_{\rm CP} & = (-0.235 \pm 0.064 \pm 0.055) \% \,.  
\end{align} 
If one uses (\ref{Master}) as a starting point and assumes the SM value for
$\Delta \Gamma_d / \Gamma_d$, then the new measurement can be used to extract
the following result for CP violation in mixing
\begin{equation} \label{Aslbnew}
  A_{sl}^b = (-0.496 \pm 0.153  \pm 0.072) \% \, ,
\end{equation}
which differs by $2.8\sp025 \sigma$ from the SM prediction \cite{Lenz:2011ti}.
The result in (\ref{Aslbnew}) is considerably smaller than the value $A_{sl}^{b}
= (-0.787 \pm 0.172 \pm 0.093)\%$ presented in 2011 \cite{Abazov:2011yk}. 
The reason for the
noticeable shift of the central value in $A_{sl}^b$ is that in
\cite{Abazov:2011yk} (as well as in all other previous experimental and
theoretical analyses) the contribution proportional to $\Delta \Gamma_d$
in~(\ref{Master}) was neglected.

Stronger statements can be obtained from the data in \cite{Abazov:2013uma}, if
different regions for the muon impact parameter (denoted by the index $i$) are
investigated separately instead of averaging over them, as done to get the
values in (\ref{New}). It is then possible to extract individual values for
$a_{sl}^d$, $a_{sl}^s$ and $\Delta \Gamma_d$ from the measurements of the
$a_{\rm CP}^i$ and $A_{\rm CP}^i$. One finds \cite{Abazov:2013uma}
\begin{align}
  \label{eq:D0:combined:measurement}
  a_{sl}^d & = (-0.62 \pm 0.43) \% \,, &
  a_{sl}^s & = (-0.82 \pm 0.99) \% \,, &
  \frac{\Delta \Gamma_d}{\Gamma_d} & = (0.50 \pm 1.38) \% \,.
\end{align}
The result differs from the combined SM expectation for the three observables by
$3.0\sp025 \sigma$. If one instead assumes that the semi-leptonic asymmetries
$a_{sl}^d$ and $a_{sl}^s$ are given by their SM values, then the decay rate
difference $\Delta \Gamma_d$ measured by \cite{Abazov:2013uma} using
(\ref{Master}) is
\begin{align} \label{eq:D0new}
  \frac{\Delta \Gamma_d}{\Gamma_d} & = (2.63 \pm 0.66) \% \; ,
      \end{align}
which differs by $3.3\sp025 \sigma$ from the SM prediction.

All in all, the new D\O{} measurements still differ from SM expectations at the
level of~$3\sp025 \sigma$ and part of this tension could be related to an
anomalous enhancement of the decay rate difference $\Delta \Gamma_d$. This
raises the question to what extent $\Delta \Gamma_d$ can be enhanced by beyond
the SM effects, without violating other experimental constraints. In fact, this
is an interesting question in its own right.  While potential new-physics
contributions to the related quantity $\Delta \Gamma_s$ have been studied in
detail (see e.g.~\cite{Dighe:2007gt, Dighe:2010nj, Bauer:2010dga, Bai:2010kf,
  Oh:2010vc, Alok:2010ij, Bobeth:2011st, Goertz:2011nx}) and turned out to be
strongly constrained by different flavour observables --- at most enhancements
of $35 \%$ are allowed experimentally --- possible new-physics effects in
$\Delta\Gamma_d$ have received much less attention. An exception is the
article~\cite{Gershon:2010wx} which emphasises that a precision measurement of
$\Delta \Gamma_d$ would provide an interesting window to new physics.

The main goal of this paper is to close the aforementioned gap by performing
model-independent studies of two types of new-physics contributions to
$\Delta\Gamma_d$ that could in principle be large. As a first possibility we
consider new-physics contributions due to current-current operators $(\bar{d}
p)(\bar p^{\sp025 \prime} b)$ with $p, p^\prime = u,c$, allowing for
flavour-dependent and complex Wilson coefficients. We study carefully the
experimental constraints on each of the coefficients that arise from hadronic
two-body decays such as $B \to \pi \pi,\, \rho\pi,\, \rho\rho,\, D^\ast \pi$,
the inclusive $B\to X_d\gamma$ decay and the dimension-eight contributions to
$\sin \left (2 \beta \right)$ \cite{Boos:2004xp} as extracted from $B \to J/\psi
K_S$.  Our analysis shows that large deviations in $\Delta\Gamma_d$ and
$a_{sl}^d$ are currently not ruled out, if they are associated to the
current-current operator involving two charm quarks.  We emphasise that our
general model-independent framework covers the case of violations of the
unitarity of the Cabibbo-Kobayashi-Maskawa (CKM) matrix.  As a second
possibility we analyse the constraints on new-physics contributions from
operators of the form $(\bar{d}b)(\bar\tau\tau)$. We show that since the
existing constraints imposed by tree-level and loop-level mediated
$B$-meson decays such as $B \to \tau^+ \tau^-$ and $B^+ \to \pi^+ \mu^+
\mu^-$ are quite loose, sizable modifications of $\Delta \Gamma_d$ and
$a_{sl}^d$ are possible also in this case, in particular if they arise from
vector operators. From a purely phenomenological point of view it thus seems
much easier to postulate absorptive new physics in the $B_d$-meson system rather
than in the $B_s$-meson system.

Our paper is organised as follows. In Section~\ref{sec:mix} we briefly set our
notation and collect the experimental values and SM predictions for the mixing
quantities. In Section~\ref{sec:DGSM} we illustrate the principle differences
between $\Delta \Gamma_d$ and $\Delta \Gamma_s$. We turn to the aforementioned
model-independent new-physics studies in Sections~\ref{sec:DGNPcc}
and~\ref{sec:DGNPtautau}, while Section~\ref{sec:conclusion} contains our
conclusions. In Appendix~\ref{app:DG} we give the SM result for
$\Delta\Gamma_d$, including a detailed breakdown of theoretical
uncertainties. The input values employed in our numerical calculations are
summarised in Appendix~\ref{app:numeric:input}.
%
%
\section{Mixing formalism}
\label{sec:mix}

Mixing phenomena in the $B_q$-meson system, with $q=d,s$, are related to the
off-diagonal elements of the complex mass matrix $M_{12}^q$ and decay rate
matrix $\Gamma_{12}^q$. We choose the three physical mixing observables as the
mass difference $\Delta M_q$, the width difference $\Delta \Gamma_q$ and the
flavour specific (or semi-leptonic) CP asymmetries $a_{sl}^q$.  The general
expressions of these observables are
\begin{align}
  \Delta M_q & 
  = 2 \left | M_{12}^q \right | \,, &
  \Delta \Gamma_q & 
  = 2 \left |\Gamma_{12}^q \right | \cos \phi_q \,, &  
  a_{sl}^q & 
  = \left| \frac{\Gamma_{12}^q}{M_{12}^q} \right| \sin \phi_q \,, 
\end{align}
with the mixing phase $\phi_q = \arg(- M_{12}^q/\Gamma_{12}^q)$. The above
equations are valid up to corrections of 
${\cal O} (1/8 \hspace{0.5mm} |\Gamma_{12}^q/M_{12}^q|^2 \hspace{0.5mm} 
\sin^2 \phi_q)$, which is smaller than $10^{-7}$ in the SM for the 
$B_q$-meson systems~\cite{Lenz:2011ti}. 

To study new-physics effects it is convenient to parameterise the general
expressions for the mixing matrices in such a way that the SM contributions
are factored out.  We employ the notation introduced in \cite{Lenz:2006hd,Lenz:2011zz}
and write
\begin{equation}
\begin{aligned}
\label{eq:M12new}
  M_{12}^q & = M_{12}^{q,\,{\rm SM}} \, \Delta_q \,, \qquad &
  \Delta_q & = |\Delta_q| \, e^{i \phi_q^\Delta} \,, 
\\[2mm]
  \Gamma_{12}^q & = \Gamma_{12}^{q,\,{\rm SM}}  \,  \tilde{\Delta}_q \,, \qquad &
  \tilde{\Delta}_q & = |\tilde{\Delta}_q| \, e^{-i \tilde{\phi}_q^\Delta} \,.
\end{aligned}
\end{equation}
The observables are then modified with respect to their SM predictions according to
\begin{equation}
 \label{eq:GammaNP}
\begin{aligned}
  \frac{\Delta M_q}{\Delta M_q^{\rm SM}} &  
  = |\Delta_q | \,, \hskip0.9cm &
  \frac{\Delta \Gamma_q}{\Delta \Gamma_q^{\rm SM}} &
  = | \tilde{\Delta}_q | \, \frac{\cos \phi_q}{\cos \phi_q^{\rm SM}} \,, \hskip0.9cm &
  \frac{a_{sl}^q}{a_{sl}^{q,\,{\rm SM}}} & 
  = \frac{|\tilde{\Delta}_q|}{|\Delta_q|}  \, 
    \frac{\sin \phi_q}{\sin \phi_q^{\rm SM}} \,,
\end{aligned}
\end{equation}
where the mixing phase is given by 
\begin{equation}
  \phi_q =  \phi_q^{\rm SM} + \phi_q^\Delta+\tilde{\phi}_q^\Delta \,.
  \label{NPphasemix}
\end{equation}

\begin{table}[t!]
\begin{center}
\renewcommand{\arraystretch}{1.1}
\begin{tabular}{|cl|ccc|ccc|}
\hline
  \multicolumn{2}{|c|}{Quantity}
& \multicolumn{3}{c|}{$q=d$}
& \multicolumn{3}{c|}{$q=s$}
\\
\hline \hline
  $\Delta M_q$
& SM
& $0.543 \pm 0.091$
& \cite{Lenz:2011ti} &
& $17.3 \pm 2.6$
& \cite{Lenz:2011ti} &
\\
\cline{2-8}
  $[\mbox{ps}^{-1}]$
& exp.
& $0.507 \pm 0.004$
& \cite{Amhis:2012bh} & HFAG
& $17.69 \pm 0.08$
& \cite{Amhis:2012bh} &  HFAG
\\
\hline\hline
  $\Delta \Gamma_q/\Gamma_q$
& SM
& $0.42 \pm 0.08$
& \cite{Lenz:2011ti} &
& ---
&  &
\\
\cline{2-8}
  $[\%]$
& exp.
& $1.5 \pm 1.8$
&  \cite{Amhis:2012bh} & HFAG
& ---
&&
\\
  $[\%]$
& 
& $0.5 \pm 1.38$
&  \cite{Abazov:2013uma} & D\O{}
& ---
&&
\\
  $[\%]$
& 
& $-4.4 \pm 2.7 \,\,$
&  \cite{Aaij:2014owa} & LHCb
& ---
&&
\\
\hline\hline
  $\Delta \Gamma_q$
& SM
& $0.0029 \pm 0.0007$ 
& see (\ref{DGdSM}) & 
& $0.087 \pm 0.021$
& \cite{Lenz:2011ti} &
\\
\cline{2-8}
  $[\mbox{ps}^{-1}]$
& exp.
& $0.0059\pm 0.0079$
& \cite{Amhis:2012bh,Abazov:2013uma} &
& $0.081 \pm 0.011$
& \cite{Amhis:2012bh} &  HFAG
\\
\hline\hline
  $a_{sl}^q$
& SM
& $-4.1 \pm 0.6$
& \cite{Lenz:2011ti} &
& $1.9 \pm 0.3$
& \cite{Lenz:2011ti} &
\\
\cline{2-8}
  $[10^{-4}]$
& exp.
& $6 \pm 17 ^{+38}_{-23}$
& \cite{Lees:2013sua} & BaBar
& $-6   \pm 50 \pm 36$
& \cite{Aaij:2013gta} & LHCb
\\
& 
& $68 \pm 45 \pm 14$
& \cite{Abazov:2012hha} & D\O{}
& $-112 \pm 74 \pm 17$
& \cite{Abazov:2012zz}  & D\O{}
\\
\hline\hline
  $\phi_q$
& SM
& $-0.085 \pm 0.025$
& \cite{Lenz:2011ti}&
& $0.0042 \pm 0.0013$
& \cite{Lenz:2011ti} &
\\
\cline{2-8}
& exp.
& ---
& &
& $\left[ -1.04, +0.96 \right]$ ${}^{\ast)}$
& \cite{Aaij:2013gta} & LHCb
\\
& 
& ---
& &
& $\left[ -1.34,  -0.58 \right]$ ${}^{\ast)}$
& \cite{Abazov:2012zz} & D\O{}
\\
\hline\hline
  $\beta$, $-2\beta_s$
& SM
& $0.446^{+0.016}_{-0.037}$
& \cite{Charles:2004jd} &
& $-0.03676^{+0.00128}_{-0.00144}$
& \cite{Charles:2004jd} &
\\
\cline{2-8}
& exp.
& $0.374 \pm 0.014$
& \cite{Amhis:2012bh} &HFAG
& $0.04^{ +0.10}_{ -0.13}$
& \cite{Amhis:2012bh} & HFAG 
\\
&
& ---
& &
& $0.01 \pm 0.07 \pm 0.01$
& \cite{Aaij:2013oba} & LHCb
\\
\hline
\end{tabular}
\renewcommand{\arraystretch}{1.0}
\caption{Collection of SM predictions and measurements (exp.) of
  $B_q$-meson mixing observables for $q = d, s$. 
  ${}^{\ast)}$As obtained from the combination of measurements of $\Delta M_s$,
  $\Delta \Gamma_s$ and $a_{sl}^s$ in the table. See text for more details.
}
\label{tab:mixing}
\end{center}
\end{table}

We now compare the SM results with the experimental measurements,
contrasting the situations in the $B_d$-meson and $B_s$-meson systems. For this
purpose we have collected the SM results of the mixing observables and the
corresponding experimental measurements in Table \ref{tab:mixing}. Concerning
$\Delta M_q$, the central values agree very nicely, but the experimental errors
are much smaller than the theoretical ones, which are dominated by hadronic
uncertainties. The experimental numbers are the current world averages obtained
by the HFAG collaboration \cite{Amhis:2012bh} and incorporate various
measurements performed at experiments from ALEPH to LHCb that are all consistent with each other.

In the case of the width difference, the SM prediction of $\Delta \Gamma_s$ agrees quite well with experimental world average
\cite{Amhis:2012bh} that combines measurements from LHCb \cite{Aaij:2013oba}, 
ATLAS \cite{Aad:2012kba}, CDF \cite{Aaltonen:2012ie} and D\O{} \cite{Abazov:2011ry}.
The very good agreement between theory and experiment shows that the Heavy Quark
Expansion (HQE) used to calculate $\Delta \Gamma_s$ works to an accuracy of
${\cal O}( 25\%)$ or better. The SM prediction for $\Delta\Gamma_d$ itself was
not explicitly given in \cite{Lenz:2011ti} but can be extracted from the
numerical code used in that work. For this reason, we give a short description
of the components that go into it and also a detailed analysis of different
sources of uncertainties in Appendix~\ref{app:DG}.  While the theory treatment
of $\Delta\Gamma_d$ is in complete analogy to $\Delta \Gamma_s$, experimental
measurements are much more challenging due to its small size, which follows
from the hierarchy of the CKM matrix elements involved.  The bounds on 
$\Delta \Gamma_d$ given in Table~\ref{tab:mixing} stem from the latest HFAG
average and two more recent investigations from D\O{}~\cite{Abazov:2013uma}
and LHCb \cite{Aaij:2014owa}. The measurements of $\Delta\Gamma_d$ are still subject
to large errors and significant deviations from the SM calculation are not yet
ruled out. We return to this in a moment.

The experimental situation for the semi-leptonic CP asymmetries is more
complicated, because the current HFAG values $a_{sl}^d = (7 \pm 27) \cdot
10^{-4}$ and $ a_{sl}^s = (-171 \pm 55) \cdot 10^{-4}$~\cite{Amhis:2012bh} are
based on the traditional interpretation (\ref{Masterold}) of the dimuon
asymmetry through CP violation in mixing alone, whereas the correct
interpretation seems to be the one given in (\ref{Master}).  We therefore quote
in Table \ref{tab:mixing} separately the direct measurements in semi-leptonic
decays from BaBar \cite{Lees:2013sua}, D\O{} \cite{Abazov:2012hha,
  Abazov:2012zz} and LHCb \cite{Aaij:2013gta}. These results for semi-leptonic
CP asymmetries are now complemented by values based on the interpretation
(\ref{Master}) of the dimuon asymmetry given in
(\ref{eq:D0:combined:measurement}).  Obviously, the experimental measurements of
the semi-leptonic asymmetries do little to restrict possible new-physics
contributions.

Alternatively, one might also compare the SM predictions for the mixing phases
with experimental constraints via the relation $\phi_q = \tan^{-1}
\left(a_{sl}^q\, \Delta M_q/\Delta \Gamma_q \right)$.  Because of the large
uncertainty in $\Delta \Gamma_d$, the experimental numbers do not provide
any stringent constraint on $\phi_d$ at present.\footnote{If one takes the $68\%$ CL
  range of the D\O{} value for the semi-leptonic asymmetry, then small negative
  values of $\phi_d$ are excluded.}  For the phase $\phi_s$ the $68\%$ confidence
  level (CL) ranges are given in Table \ref{tab:mixing}.

In addition to the three mixing observables $\Delta M_q$, $\Delta \Gamma_q$
and $a_{sl}^q$, the phase of $M_{12}^q$ affects time-dependent CP asymmetries
of neutral $B_q$-meson decays. Neglecting CP violation in mixing and
corrections of ${\cal O} (|\Gamma_{12}^q/M_{12}^q|^2)$, the interference
between mixing and decay, 
\begin{align}
  \label{eq:def:time-dep:CPasy}
  \frac{{\rm Br}\left(\bar{B}_q(t)\to f \right) - {\rm Br} \left(B_q(t)\to f \right)}
       {{\rm Br}\left(\bar{B}_q(t)\to f \right) + {\rm Br} \left(B_q(t)\to f \right)}
  & =  
  S_f \sin \left(\Delta M_q \, t \right )-C_f \cos \left (\Delta M_q \, t \right ) \,, 
\end{align}
gives rise to the direct and the mixing-induced CP asymmetries $C_f$ and $S_f$, 
respectively. Within these approximations, especially $S_f$ is sensitive to the
new physics phase $\phi_q^\Delta$ of $M_{12}^q$
\begin{align}
  \label{eq:Sdef}
  S_f & 
  =  \frac{2\,{\rm Im} \, 
  \left[ e^{-i\, \left(2 \arg \left (V_{tq}^\ast V_{tb}^{} \right ) 
        + \phi_q^\Delta\right)}\rho_f \right]}
  {1+|\rho_f|^2} \,,
\end{align}
where $\rho_f \equiv \bar{\cal A}_f/{\cal A}_f$ and $\bar{\cal A}_f$ $({\cal
  A}_f)$ is the hadronic amplitude of the $\bar{B}_q$ ($B_q$)-meson decay. In
cases where the hadronic amplitude is dominated by a single weak phase, such as 
in the SM, $S_f$ can be cleanly related to the corresponding CKM parameters. This
is for instance the case for decays which are dominated by the tree
$b\to c\bar{c}s$ transition or the gluonic penguin $b\to q\bar{q} s$~($q = u,d,s$)
transitions.

The two well-known examples of $b\to c\bar{c}s$  mediated tree decays are
$B_d \to J / \psi K_S$ and $B_s \to J / \psi \phi$, both of which measure the
relative phase between the hadronic decay amplitude and $M_{12}^q$. 
In this case one has 
\begin{align}
  \label{eq:NPphaseint}
  S_f = -\sin\left(\pm 2 \beta^{f}_{q}\right) & 
  = -\sin \left( \pm 2\beta_{q}  + \delta_q^{\Delta} 
   + \delta_q^{\rm peng,\, NP} 
        + \delta_q^{b\to c\bar{c}s}\right) \,, 
\end{align} 
with $+$ for $q = d$ and $-$ for $q = s$. The dominant SM amplitude
contributes $\beta_d = \beta = \arg \left(-V_{cb}^\ast V_{cd}^{}/V_{tb}^\ast
  V_{td}^{} \right)$ and $\beta_s = \arg \left( -V_{tb}^\ast V_{ts}^{}/
  V_{cb}^\ast V_{cs}^{} \right)$, respectively. New-physics effects can
change either the phase $\phi_q^\Delta$ of $M_{12}^q$ $\big($see
(\ref{eq:M12new})$\big)$, whose contribution is denoted by $\delta^\Delta_q$,
the phase of ${\cal A}_f$ or both. The new-physics contributions to the hadronic
decay amplitude can be further subdivided into the ones from penguin
contributions $\delta_q^{\rm peng,\, NP}$ and the new-physics phase
$\delta_q^{b\to c\bar{c}s}$ appearing in the $b\to c\bar{c} {s}$ tree-level
amplitude. The SM penguin contributions are expected to be quite small
(see~e.g.~\cite{Faller:2008gt}). This general parametrisation implies that the
$\delta_q^{i}$ are actually dependent on both weak and strong phases
specific to the final state $f$. In the current work we are only interested in
new physics that does not introduce visible effects in the penguin sector and
furthermore consider only interactions that affect the $b\to c\bar{c}d$
tree-level decay, but not $b\to c\bar{c} {s}$. In consequence, we set
$\delta_q^{\rm peng,\, NP}$ and $\delta_q^{b\to c\bar{c}s}$ to zero
in~(\ref{eq:NPphaseint}).

The lesson to learn from the above comparisons is that with the exception of
$\Delta M_d$, the mixing observables in the $B_d$-meson system are rather
loosely constrained experimentally.  There is thus in principle plenty of room
for beyond the SM contributions to $\Delta\Gamma_d$ and $a_{sl}^d$, which depend
besides $M_{12}^d$ also strongly on the non-standard effects in
$\Gamma_{12}^d$. Within the SM, $\Gamma_{12}^d$ is determined by $\Delta B = 1$
operators of dimension six, predominantly those arising from tree-level
$W$-boson exchange $b\to f_1\bar{f}_2 d$ with $f_i = u,c$. Beyond the SM, other
contributions are conceivable, however the masses of $f_i$ should allow for the
formation of intermediate on-shell states in order to affect
$\Gamma_{12}^d$. Below we will study constraints on such absorptive
contributions within a model-independent, effective field theory framework for
the two cases $f_i = u,c$ and $f_1=f_2=\tau$.

A further important point to keep in mind is that $\Gamma^d_{12}$ is related to
a product of two $\Delta B=1$ operators, while $M_{12}^d$ is determined from 
$\Delta B = 2$ operators that are obtained by integrating out high-virtuality
particles. In the effective field theory approach the two types of operators are
independent of one another. Motivated by this observation we will study the case
where new physics manifests itself to first approximation only in terms of 
$\Delta B = 1$ interactions of the form $b\to f_1\bar{f}_2 d$, which change 
$\Gamma^d_{12}$ (i.e. $\tilde{\Delta}_d \neq 1$), while the leading dimension-six 
$\Delta B=2$ operators are assumed to remain SM-like. 
In this way we can get an idea of how large the maximum effects in $\Gamma_{12}^d$ can 
be. Of course, in explicit new-physics models the Wilson coefficients
of both types of operators will generically be modified, which can lead to
correlated effects in $M_{12}^d$ and $\Gamma_{12}^d$, thereby reducing the
possible shifts in $\Gamma_{12}^d$. Studying such correlations requires to
specify a concrete model, which  goes beyond the scope of this work. However,
even in our approach, the dimension-six $\Delta B = 1$ operators give rise to
dimension-eight $\Delta B = 2$ operators due to operator mixing, and consequently,
induce also new-physics contributions to $M_{12}^d$ (i.e. $\Delta_q \neq 1$). 
We will study these effects  in Section~\ref{sec:DGNPcc} and show that they are phenomenologically
harmless for the effective interactions considered in our analysis. 

%
%
\section{Comparison of $\bm{\Delta \Gamma_d}$ and $\bm{\Delta \Gamma_s}$}
\label{sec:DGSM}

The first important observation is that $\Delta \Gamma_d$ is triggered by the
CKM-suppressed decay $b \to c \bar{c} d$, whose inclusive branching ratios reads
$(1.31 \pm 0.07) \%$ 
(based on the numerical evaluation in \cite{Krinner:2013cja}), 
while $\Delta \Gamma_s$ receives
the dominant contribution from the CKM-favoured decay $b \to c \bar{c} s$, which
has an inclusive branching ratio of $(23.7 \pm 1.3) \%$
\cite{Krinner:2013cja}. This means that a relative modification of $\Gamma(b \to
c \bar{c} d)$ by 100\% shifts the total $b$-quark decay rate 
$\Gamma_{\rm tot}$\footnote{We do not distinguish here between the total $B$-meson decay rates
$\Gamma_{B_d}$, $\Gamma_{B_s}$ and $\Gamma_{B^+}$  and 
the total $b$-quark decay rate $\Gamma_{\rm tot}$, because the measured differences 
are smaller than the current theoretical uncertainites.}
by around 1\% only, while a 100\% variation in $\Gamma(b \to c \bar{c} s)$
results in an effect of roughly 25\% in the same observable. Large enhancement
of the $b \to c \bar{c} d$ decay rate can therefore be hidden in the hadronic
uncertainties of $\Gamma_{\rm tot}$, while this is not possible in the case of
the $b \to c \bar{c} s$ decay rate.

Second, the CKM structure of $\Gamma_{12}^d$ and $\Gamma_{12}^s$ are notably
different within the SM. Separating $\Gamma_{12}^q$ into the individual
contributions with only internal charm quarks ($\Gamma_{12}^{cc,q}$) or up
quarks ($\Gamma_{12}^{uu,q}$) and one internal charm quark and one up quark
($\Gamma_{12}^{cu,q}$), one can write the SM contribution to $\Gamma_{12}^q$ as
follows
\begin{align} 
  \label{eq:Gamma12qformula}
  \Gamma_{12}^{q,{\rm SM}} & 
  = - \left[\left( \lambda_c^q \right)^2 \Gamma_{12}^{cc,q}
          + 2\, \lambda_c^q \lambda_u^q\, \Gamma_{12}^{cu,q}
          + \left( \lambda_u^q \right)^2 \Gamma_{12}^{uu,q} \right] \, ,
\end{align}
with $\lambda^q_p \equiv V_{pq}^\ast V_{pb}^{}$ and \cite{Lenz:2011ti} 
\beq \label{eq:Gamma12pq}
\Gamma_{12}^{cc,d}  = 18.85 \, \mbox{ps}^{-1} \, , \qquad 
  \Gamma_{12}^{cu,d}  = 20.72\, \mbox{ps}^{-1} \, , \qquad
     \Gamma_{12}^{uu,d}  = 22.52 \, \mbox{ps}^{-1} \,.
\eeq
The results for the $\Gamma_{12}^{pp^\prime,s}$ coefficients relevant in the
$B_s$-meson system are obtained from the latter numbers by a simple rescaling
with the factor $(f_{B_s} M_{B_s})^2/(f_{B_d} M_{B_d})^2 = 1.48$. Notice that
the three values in (\ref{eq:Gamma12pq}) are quite similar, which implies that
phase-space effects are not very pronounced.

Combining the formulae (\ref{eq:Gamma12qformula}) and (\ref{eq:Gamma12pq}), we
obtain the following numerical expressions
\beq 
  \label{eq:G12split}
\begin{split}
  \Gamma_{12}^{d,{\rm SM}} & 
    = -\big [ ( 1.60 + 0.00 \hspace{0.5mm} i )  
            + (-0.50 + 1.37 \hspace{0.5mm} i ) 
            + (-0.25 - 0.21 \hspace{0.5mm} i ) \big ] \cdot 10^{-3} \, {\rm ps}^{-1} 
\\[2mm] 
  & = -\big [ 0.85 + 1.16 \hspace{0.5mm} i \big]  \cdot 10^{-3} \, {\rm ps}^{-1} \,, 
\\[3mm]
  \Gamma_{12}^{s,{\rm SM}} & 
    = -\big [ (42.81 + 0.00 \hspace{0.5mm} i )
            + ( 0.72 - 1.97 \hspace{0.5mm} i )
            + (-0.02 - 0.02 \hspace{0.5mm} i ) \big ] \cdot 10^{-3} \, {\rm ps}^{-1} 
\\[2mm]
  & = -\big [ 43.51 - 1.98 \hspace{0.5mm} i \big ]  \cdot 10^{-3} \, {\rm ps}^{-1} \,.
\end{split}
\eeq 
From the first line of the result for $\Gamma_{12}^{d,{\rm SM}}$, we see
that in the $B_d$-meson system there is a partial cancellation between the
individual contributions, because the two relevant CKM factors are of similar
size in this case, i.e.~$|\lambda_c^d| \simeq |\lambda_u^d| = {\cal O}
(\lambda^3)$ with $\lambda \simeq 0.22$ denoting the Cabibbo angle.  In the case
of $\Gamma_{12}^s$, on the other hand, the result is fully dominated by the
contribution due to $\Gamma_{12}^{cc,s}$, since $|\lambda_c^s| = {\cal O}
(\lambda^2)$ while $|\lambda_u^s| = {\cal O} (\lambda^4)$. The observed partial
cancellation leads again to the feature that a modification in $b \to c \bar{c}
d$ will have a much larger effect in $\Gamma_{12}^d$, compared to the effect of
a similar modification in $b \to c \bar{c} s$ in $\Gamma_{12}^s$. For instance,
a $100\%$ shift in $\Gamma_{12}^{cc,d}$ leads to an almost $300\%$ modification
of $\Gamma_{12}^d$, while a $100\%$ change of $\Gamma_{12}^{cc,s}$ results in a
$100\%$ shift of $\Gamma_{12}^s$ only.

Another way of looking at the mixing systems is to investigate the ratio
$\Gamma_{12}^q/M_{12}^q$. Using the unitarity of the CKM matrix,
i.e.~$\lambda_u^q + \lambda_c^q + \lambda_t^q =0$, we find the SM expression
\beq 
  \left ( \frac{\Gamma_{12}^q}{M_{12}^q} \right )_{\rm SM} 
  = \left [ (-51 \pm 10) 
          + \frac{\lambda_u^q}{\lambda_t^q} \, (10 \pm 2) 
          + \left(\frac{\lambda_u^q}{\lambda_t^q}\right)^2 \, (0.16 \pm 0.03)
    \right]\cdot 10^{-4}\, ,
  \label{MasterI}
\eeq
where the numerical coefficients are identical for the $B_d$-meson and
$B_s$-meson system within errors. The relevant CKM factors are $\lambda_u^d/
\lambda_t^d = -0.033 - 0.439 \hspace{0.5mm} i$ and $\lambda_u^s/ \lambda_t^s =
-0.008 + 0.021\sp025  i$. It follows that the real part of
$(\Gamma_{12}^q/M_{12}^q)_{\rm SM}$ and thus $(\Delta \Gamma_q / \Delta
M_q)_{\rm SM}$ is dominated by the first term in the
square bracket of~(\ref{MasterI}), which encodes the contribution to
$\Gamma_{12}^{q,{\rm SM}}$ involving charm quarks only. The situation is quite
different for the imaginary parts of $(\Gamma_{12}^q/M_{12}^q)_{\rm SM}$ that
arise from the second and third term and determine the size of $a_{sl}^{q, {\rm
  SM}}$. The fact that in the SM the semi-leptonic CP
asymmetry in the $B_d$-meson sector is about~20 times larger than the one in the
$B_s$-meson system and has opposite sign is hence a simple consequence of ${\rm
  Im} \left (\lambda_u^d/ \lambda_t^d \right ) \simeq - 20 \hspace{0.75mm} {\rm
  Im} \left (\lambda_u^s/ \lambda_t^s \right )$.

The structure of $\Gamma_{12}^{q,{\rm SM}}$ and $\left ( \Gamma_{12}^q/M_{12}^q
\right )_{\rm SM}$ also allows one to draw some general conclusions on how new
physics can modify $\Delta \Gamma_{d,s}$ and $a_{sl}^{d,s}$. Consider for
instance the violation of CKM unitarity $\lambda_u^q + \lambda_c^q + \lambda_t^q
= \Delta_{\rm CKM}^q$, a property known from beyond the SM
scenarios~(see~e.g.~\cite{delAguila:2000rc,Casagrande:2008hr,Eberhardt:2010bm})
in which heavy fermions mix with the SM quarks and/or new charged gauge bosons
mix with the $W$ boson.\footnote{See also \cite{Botella:2014qya} for a recent
  discussion of a similar point.} In such models the relation (\ref{MasterI}) would
receive a shift that can be approximated by
\beq 
  \label{eq:DeltaGoM}
  \Delta \left ( \frac{\Gamma_{12}^q}{M_{12}^q} \right ) \simeq    
   (-51 \pm 10) \left [ \left ( 1 - \frac{\Delta_{\rm CKM}^q}{\lambda_t^q} \right )^2 -1 \right ] \cdot 10^{-4}\, .
\eeq
Given our imperfect knowledge of some of the elements of the CKM matrix,
deviations of the form $\Delta_{\rm CKM}^d \simeq \Delta_{\rm CKM}^s = {\cal O}
(\lambda^3)$ are not excluded phenomenologically. From (\ref{eq:DeltaGoM}) we
then see that such a pattern of CKM unitarity violation can lead to a relative
enhancement of $|\Gamma_{12}^d/M_{12}^d|$ by up to $300\%$, while in the case of
$|\Gamma_{12}^s/M_{12}^s|$ the relative shifts can be $50\%$ at most.  Depending
on the phase of $\Delta_{\rm CKM}^d/\lambda_t^d$ the new contribution in
(\ref{eq:DeltaGoM}) could hence affect $\Delta \Gamma_d$ and $a_{sl}^d$ in a
significant way, while leaving $\Delta M_d$, $\Delta M_s$, $\Delta \Gamma_s$ and
$a_{sl}^s$ unchanged within hadronic uncertainties. In fact, in the next two
sections we will see that it is possible to find certain effective interactions
that support the general arguments presented above.

%
%
\section{New physics in $\bm{\Delta \Gamma_d}$: current-current operators }
\label{sec:DGNPcc}

In the following we derive model-independent bounds on the Wilson coefficients
of so-called current-current operators.  We write the part of the effective
weak Hamiltonian involving these operators as
\begin{equation} 
  \label{eq:HeffQ1Q2} 
  {\cal H}^{\rm current}_{\rm eff} = \frac{4 G_F}{\sqrt{2}} 
  \sum_{p,p^\prime=u,c}  V_{pd}^\ast 
\sp025  V_{p^\prime b}^{}  \sum_{i=1,2} 
  C_i^{pp^\prime}(\mu) \sp025  Q_i^{pp^\prime}
  + {\rm h.c.} \,, 
\end{equation}
with $G_F$ the Fermi constant. The current-current operators are then defined
as
\begin{align}
  \label{eq:Q1Q2}
  Q_1^{pp^\prime} &
  = \big( \bar d^{\, \alpha} \hspace{0.5mm} \gamma^\mu P_L \sp025  p^{\beta} \big) 
    \big( \bar p^{\, \prime \beta} \hspace{0.0mm} \gamma_\mu P_L \hspace{0.0mm} b^\alpha \big) \,, &
  Q_2^{pp^\prime} &
  = \big( \bar d \hspace{0.0mm} \gamma^\mu P_L \sp025  p \big) 
    \big( \bar p^{\, \prime} \hspace{0.0mm} \gamma_\mu P_L \hspace{0.0mm}  b \big) \,,
\end{align}
where $P_L = (1-\gamma_5)/2$ projects onto left-handed fields and $\alpha,\beta$
denote colour indices.

In the SM, the coefficients $C_i^{pp^\prime}$ are real and depend neither on
the quark content $pp^\prime$ nor on whether the transition is $b\to s$ or
$b\to d$.  On the other hand, a generic new-physics model will give rise to
different contributions to each non-leptonic decay channel and one must consider
carefully the constraints on the (complex) coefficients $C_i^{pp^\prime}$
individually. While some ingredients needed for such a study have been mentioned
previously in the literature, see for instance \cite{Bauer:2010dga}, it has yet
to be carried out in any detail.  The goal of this section is to fill this gap
by deriving bounds on the $C_i^{uu}$, $C_i^{uc}$ and $C_i^{cc}$ coefficients,
which multiply the operators governing $b\to u \bar{u} d$, $b\to c \bar{u} d$
and $b\to c\bar{c} d$ transitions, respectively.\footnote{The coefficient
  $C_i^{cu}$ governs $b \to u {\bar c} d$ transitions, which are severely CKM
  suppressed in the SM. Since currently the most stringent bounds on such a
  coefficient come from $\Delta\Gamma_d$ itself, we exclude it from the
  analysis.}  We structure this section by discussing the coefficients
$C_{1,2}^{pp^\prime}$ in turn, examining not only the constraints but also their
implications for deviations in $\Delta\Gamma_d$ from the SM expectation.  To do
so, we first write
\begin{align}
  \label{eq:Csplit}
  C_{1,2}^{ij} & = C_{1,2}^{\rm SM} + \Delta C_{1,2}^{ij} \,.
\end{align}
Then, generalising the expressions from Section~\ref{sec:DGSM}, we obtain
\begin{align}
\label{eq:dg12}
\frac{\Gamma^d_{12}}{\Gamma_{12}^{d,{\rm SM}}}-1&
=\left(0.61-0.84 \sp025  i\right) 
\bigg[\left(\Delta C_{2}^{cc}\right)^2+0.064  \sp025  
\Delta C_{2}^{cc}  \sp025   \Delta C_{1}^{cc}+2.1 \sp025   \Delta C_{2}^{cc}\nonumber\\
  & \hspace{3.2cm}-0.26  \sp025   \Delta C_{1}^{cc}+0.77 
\left(\Delta C_{1}^{cc}\right)^2\bigg]\nonumber\\
& \phantom{xx} - \left(0.21-0.052  \sp025   i\right)\bigg[\left(\Delta C_{2}^{uu}\right)^2
+0.35  \sp025   \Delta C_{1}^{uu}  \sp025   \Delta C_{2}^{uu}+2.0  \sp025  \Delta C_{2}^{uu}\nonumber\\
&\hspace{3.6cm}-0.16 \sp025   \Delta C_{1}^{uu}
+1.3 \left(\Delta C_{1}^{uu}\right)^2\bigg]\nonumber\\
&\phantom{xx} +\left(0.53+0.79 \sp025    i\right)\bigg[\Delta C_{2}^{cu}
\Delta C_{2}^{uc}+ 1.05  \sp025   \Delta C_{1}^{cu}
\Delta C_{1}^{uc}\nonumber \\
& \hspace{3.4cm}
+0.11 \sp025   \big (\Delta C_{1}^{uc}\Delta C_{2}^{cu}
+\Delta C_{1}^{cu}\Delta C_{2}^{uc} \big ) \nonumber\\
& \hspace{3.4cm}
+1.0  \sp025   \big (\Delta C_{2}^{cu}+\Delta C_{2}^{uc} \big )
-0.10  \sp025   \big (\Delta C_{1}^{cu}+\Delta C_{1}^{uc} \big )\bigg] \,.
\end{align}
Here and in the remainder of the section we use a notation where the Wilson
coefficients $\Delta C_{1,2}^{pp^\prime}$ are to be evaluated at the scale $m_b = \m_b (\m_b)
\simeq 4.2 \, {\rm GeV}$ unless otherwise specified.  Given this expression, it
is straightforward to calculate the ratio $\Delta\Gamma_d/\Delta\Gamma_d^{\rm
  SM}$ using (\ref{eq:GammaNP}).

While it is the coefficients $\Delta C_{1,2}^{pp^\prime}(m_b)$ which appear in
low-energy observables such as~(\ref{eq:dg12}), it is important to keep in mind
that these are obtained from the matching coefficients at the new-physics scale
$\Lambda_{\rm NP}$ through renormalisation-group (RG) evolution.  In what
follows, we will always present bounds on the coefficients at the scale $M_W$
for convenience. The leading-logarithmic (LL) evolution connecting the
coefficients at the two scales can be written as \beq
\begin{split}
  \label{eq:C1C2ev}
  \Delta C_1^{pp^\prime}(m_b) & 
  = z_+ \hspace{0.5mm} \Delta C_1^{pp^\prime}(M_W)
  + z_- \hspace{0.5mm} \Delta C_2^{pp^\prime}(M_W) \,, 
\\[1mm]
  \Delta C_2^{pp^\prime}(m_b) &
  = z_- \hspace{0.5mm} \Delta C_1^{pp^\prime}(M_W)
  + z_+ \hspace{0.5mm} \Delta C_2^{pp^\prime}(M_W) \,,
\end{split}
\eeq
where 
\beq \label{eq:zpm}
z_{\pm}=\frac{1}{2}\left(\eta_5^{\frac{6}{23}} \pm  \eta_5^{-\frac{12}{23}} \right) \,, 
\qquad \eta_5 \equiv \frac{\alpha_s(M_W)}{\alpha_s(m_b)} \,,
\eeq
and $\alpha_s$ should be evaluated in the five-flavour theory.  Throughout our
work, in deriving the constraints on the individual coefficients from a given
observable, we work under the assumption of ``single operator dominance'' and
consider only changes in the coefficients one at a time. E.g.~to set constraints
on $\Delta C_1^{uu}(M_W)$ we fix $\Delta C_2^{uu}(M_W)=0$.

The dominant effect of the modifications of the current-current sector considered here is to 
change $\Gamma_{12}^d$ from its SM value.  However,  double insertions of $\Delta B =
1$ operators also give dimension-eight contributions to $M_{12}^q$.  It turns out that these 
are completely negligible numerically with the exception of the contributions from the $Q_{2}^{c c}$
operator.  We thus postpone a more detailed discussion of these contributions  to 
Section~\ref{sec:cc}.
%
%
\subsection{Bounds on up-up-quark operators}
\label{sec:uu}

We begin our analysis by deriving constraints on $C_{1,2}^{uu}$ from $B\to
\pi\pi,\, \rho\pi,\, \rho\rho$ decays.\footnote{Constraints on the real and imaginary parts of $C_{1,2}^{uu}$ can also be derived from studies of the isospin asymmetry in $B \to \rho \gamma$ \cite{Lyon:2013gba}. The obtained bounds would benefit greatly  from better measurements of the $B \to \rho \gamma$ decay. }  QCD factorisation provides a tool for
calculating various observables in these decays to leading power in
$\Lambda_{\rm QCD}/m_b$ \cite{Beneke:2003zv}.  The reliability of the
factorisation predictions is a subject of debate, and in the following we
consider only observables which can be argued to be under theoretical control
within this approach.

The process $B^- \to \pi^- \pi^0$ is to an excellent approximation a pure tree
decay and thus provides strong constraints on the magnitudes and phases of the
Wilson coefficients $C_{1,2}^{uu}$.  A particularly clean probe of tree
amplitudes is provided by the ratio \cite{Bjorken:1988kk}
\begin{align} 
  \label{eq:Rpipipi} 
  R_{\pi^- \pi^0} &
  = \frac{\Gamma (B^- \to \pi^- \pi^0)}
         {d\Gamma (\bar B^0 \to \pi^+ \ell^- \bar \nu_\ell)/dq^2 \big |_{q^2=0}}
  \simeq 3\pi^2 \sp025 f_\pi^2 \hspace{0.75mm} |V_{ud}|^2 \,
   \big | \alpha_1 (\pi \pi) + \alpha_2 (\pi \pi) \big |^2 \,, 
\end{align}
which by construction is free of the uncertainty related to $|V_{ub}|$ and the
$B \to \pi$ form factor $F_0^{B \to \pi} (0)$. Here $\Gamma (B^-
\to \pi^- \pi^0)$ denotes the $B^- \to \pi^- \pi^0$ rate, $d\Gamma (\bar B^0 \to
\pi^+ \ell^- \bar \nu_\ell)/dq^2 \big |_{q^2=0}$ is the semi-leptonic decay spectrum
differential in the dilepton invariant mass $q^2$, evaluated at $q^2 = 0$.
The numerical values of the pion decay constant $f_\pi$ and of $|V_{ud}|$ are
collected in~Table \ref{tab:numeric:input}.

\begin{figure}[!t]
\begin{center}
\vspace{-5mm}
  \includegraphics[height=0.425 \textwidth]{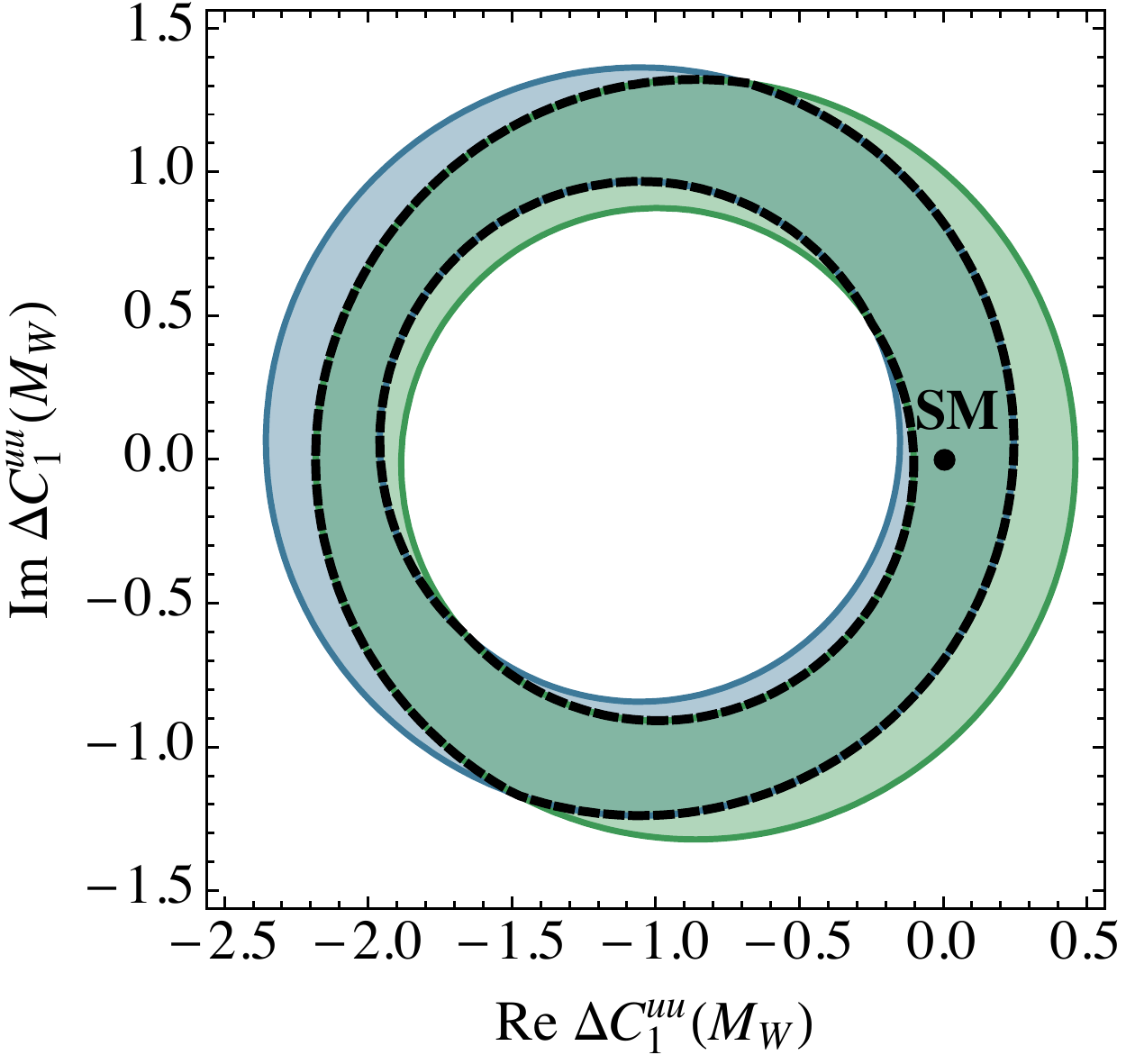} \qquad
  \includegraphics[height=0.425 \textwidth]{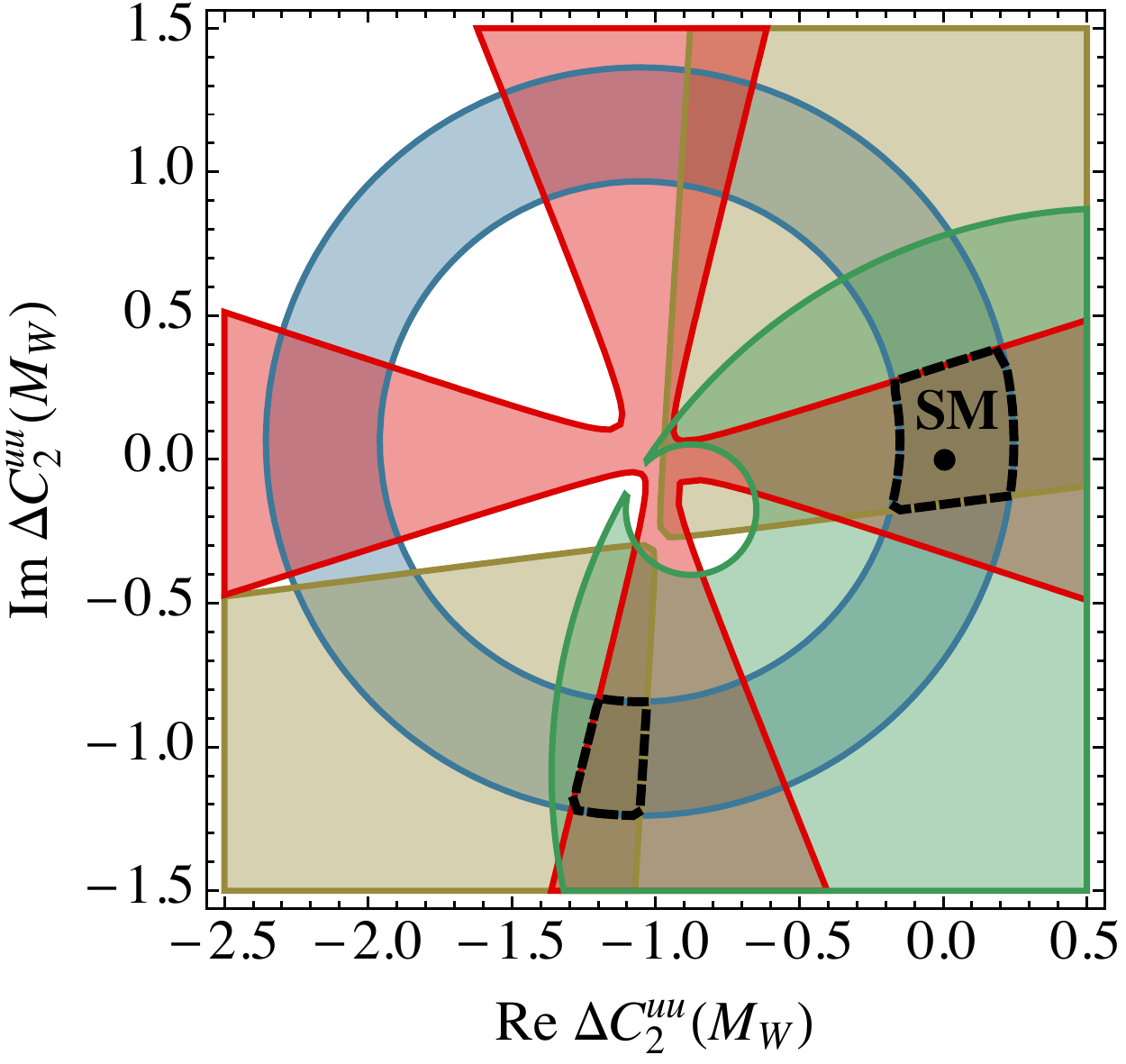} \\[4mm]
  \includegraphics[height=0.425 \textwidth]{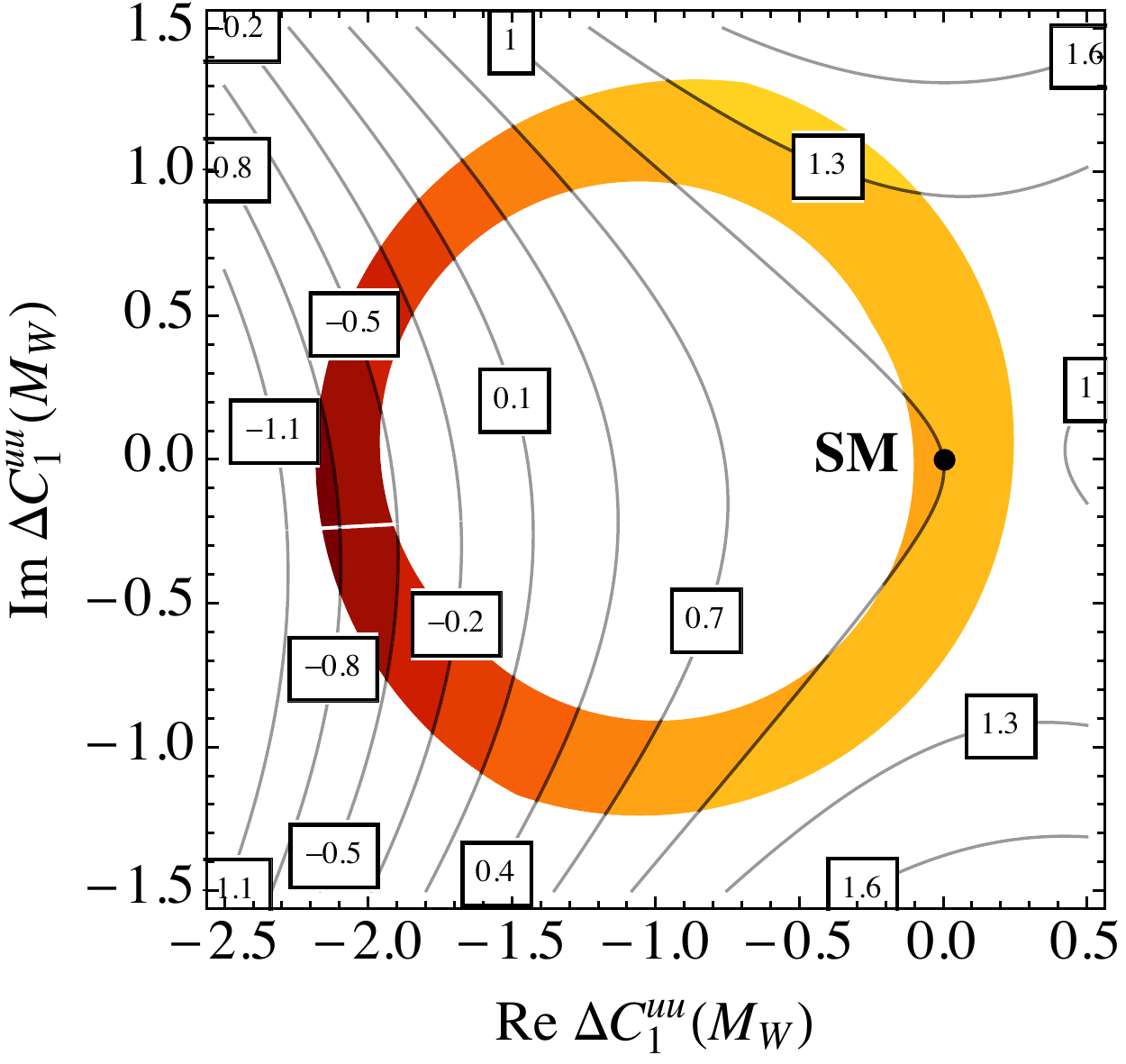} \qquad
  \includegraphics[height=0.425 \textwidth]{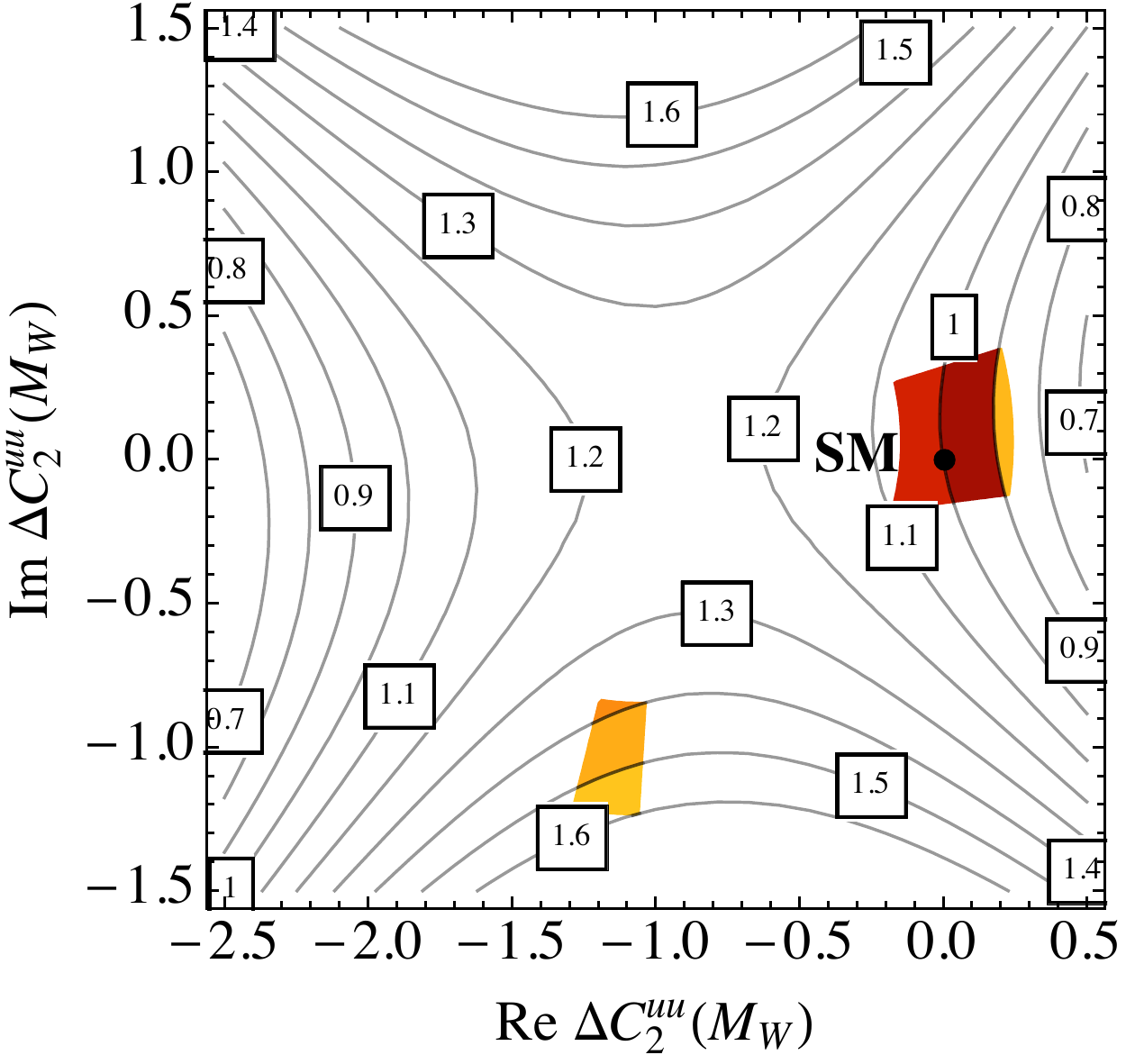}
\caption{\label{fig:DC12uuG} Upper panels: allowed parameter space in the ${\rm
    Re} \sp025 \Delta C_{1,2}^{uu} \sp025$--$\sp025 {\rm Im} \sp025
  \Delta C_{1,2}^{uu}$ planes. The blue, brown, red, and green regions are
  related to the constraints derived from $R_{\pi^- \pi^0}$, $S_{\pi^+ \pi^-}$,
  $S_{\rho\pi}$ and $R \left (\rho^- \rho^0/\rho^+ \rho^- \right )$,
  respectively. The regions enclosed by the dashed black lines represent the
  combined constraint from the different observables. Lower panels: contours of
  $\Delta\Gamma_d/\Delta\Gamma_d^{\rm SM}$ along with the combined constraints.
}
\end{center}
\end{figure}

Ignoring small electroweak penguin amplitudes the ratio $R_{\pi^- \pi^0}$
measures the magnitude of the sum of the coefficients $\alpha_{1,2}(\pi\pi)$ of
the tree amplitudes \cite{Beneke:2003zv}. These coefficients are currently known
to next-to-next-to-leading order~(NNLO) in perturbation theory
\cite{Bell:2009fm, Beneke:2009ek}.  Working to NNLO in the SM, but to leading
order~(LO) in the new-physics contributions $\Delta C_{1,2}^{uu}$ in
(\ref{eq:Csplit}), we obtain the expressions\footnote{Note that $\alpha_{1,2}$
  are interchanged with respect to the common definition in the literature
  \cite{Beneke:2003zv}, to comply with our choice of operator basis.}
\begin{equation} \label{eq:alpha1alpha2}
\begin{split}
  \alpha_1 (\pi\pi) = 0.195 - 0.101 \sp025  i + \Delta C_1^{uu}  
    + \frac{\Delta C_2^{uu}}{3} \,,
\\[2mm]
  \alpha_2 (\pi\pi) = 1.013 + 0.027 \sp025  i + \Delta C_2^{uu}  
    + \frac{\Delta C_1^{uu}}{3} \,.
\end{split}
\end{equation}
The given SM values correspond to the central values for $\alpha_{1,2}(\pi\pi)$
presented in \cite{Bell:2009fm}.  Inserting (\ref{eq:alpha1alpha2}) into
(\ref{eq:Rpipipi}), one obtains the approximation
\begin{equation}
  \label{eq:RpipipiNP}
\begin{split}
  R_{\pi^- \pi^0} \simeq \left (0.70^{+0.12}_{-0.08} \right )
  \Big[ 1+ 2.20 \hspace{0.5mm} {\rm Re} \sp025 \Delta C_+^{uu} 
         - 0.13 \hspace{0.5mm} {\rm Im} \sp025 \Delta C_+^{uu}
         + 1.21 \left | \Delta C_+^{uu} \right |^2  \Big ]  \, {\rm GeV}^2 \,,
\end{split}
\end{equation}
where $\Delta C_+^{pp^\prime} \equiv \Delta C_1^{pp^\prime} + \Delta C_2^{pp^\prime}$ and we have
restored the theoretical error of the SM prediction for $R_{\pi^- \pi^0}$ as
given in \cite{Bell:2009fm}. The corresponding experimental value is
\begin{align} 
  R_{\pi^- \pi^0} = \left (0.81 \pm 0.14 \right ) {\rm GeV}^2 \,, 
\end{align}
which has been derived in \cite{Bell:2009fm} based on the information given in
\cite{Amhis:2012bh,Ball:2006jz}.  Combining the theoretical expectation
(\ref{eq:RpipipiNP}) with the experimental determination allows to directly
constrain the magnitude and phase of $\Delta C_+^{uu}(m_b)$.  In order to turn
this constraint into individual bounds on $\Delta C_{1,2}^{uu}(M_W)$, we use
(\ref{eq:C1C2ev}) along with the assumption of single operator dominance
explained after (\ref{eq:zpm}). In Figure~\ref{fig:DC12uuG} we show the
parameter ranges (blue circular bands) that are allowed at 90\% CL
using this procedure.  We see that ${\cal O} (1)$ effects in $\Delta
C_{1,2}^{uu}(M_W)$ are allowed if they leave the magnitude $|\alpha_1 (\pi \pi)
+ \alpha_2 (\pi \pi)|$ SM-like.

A very effective way to constrain the phases of the Wilson coefficients in
addition to their magnitude is to study mixing-induced CP asymmetries $S_f$
(\ref{eq:Sdef}) in $B\to\pi\pi,\, \pi\rho,\, \rho\rho$ transitions.  On the
other hand, the direct CP asymmetries, $C_f$, are suppressed by powers of
$\alpha_s$ and/or $\Lambda_{\rm QCD}/m_b$ in QCD factorisation and are difficult
to predict quantitatively.  The mixing-induced CP asymmetries are directly
proportional to $\sin\left (2\beta + 2\gamma \right)$ 
in the SM, in the limit where the Wilson coefficients of the
QCD penguin operators are ignored.  When the Wilson coefficients $\Delta
C_{1,2}^{uu}$ are allowed to be complex, this SM relation is altered to include
the phase of the coefficients, even at LO in $\alpha_s$. The situation is more
complicated when penguin corrections are included.  However, the penguin-to-tree
ratio is of ${\cal O} (30\%)$ in $B\to \pi\pi$ and of ${\cal O}( 10\%)$ in $B\to
\rho\pi$ \cite{Beneke:2003zv}, so that the constraints given by these
observables can still be considered relatively clean.

We can evaluate the indirect asymmetries in the $B\to\pi\pi, \rho\pi$ sectors at
NLO in QCD factorisation using the formulae given in
\cite{Beneke:2003zv}.\footnote{We have also studied the indirect asymmetry
  $S_{\rho^+\rho^-}$.  The constraints are qualitatively similar to those from
  $S_{\rho\pi}$ and do little to cut down the allowed parameter space, so we
  exclude them from Figure~\ref{fig:DC12uuG} for simplicity.}  We find 
\begin{align}
 \label{eq:Spipi} 
S_{\pi^+\pi^-} \simeq -(47 \pm 28) \left [ \frac{3.48 \, {\rm Im} \, r_{\pi^+
      \pi^-} + 1.87 \, {\rm Re} \, r_{\pi^+ \pi^-}}{1 + 0.87 \left |
      r_{\pi^+ \pi^-} \right |^2 } \right ] \% \,, 
\end{align}
with 
\begin{align}
r_{\pi^+ \pi^-} = \frac{1 + \left (0.94 - 0.27 \, i \right ) \Delta
  C_{1/3}^{uu}}{1 + \left (0.88 + 0.25 \, i \right ) (\Delta
  C_{1/3}^{uu})^\ast} \,,
\end{align}
and 
\begin{equation}
 \label{eq:Srhopi}
 \begin{split} 
S_{\rho\pi}  & \equiv \frac{1}{2}\left(S_{\rho^+\pi^-} + S_{\rho^-\pi^+}\right) \\[2mm]
& \simeq (6.9 \pm 34)\, \Bigg\{
0.62 \left[ \frac{-27.3 \, {\rm Im} \, r_{\rho^-\pi^+} + 2.4 \, {\rm Re} \, r_{\rho^-\pi^+}}{1 + 1.4 \left |
      r_{\rho^-\pi^+} \right |^2 } \right ]  \\ 
&\hspace{2.4cm}+ 0.38  \left[ \frac{-31.7 \, {\rm Im} \, r_{\rho^+\pi^-} + 1.7 \, {\rm Re} \, r_{\rho^+\pi^-}}{1 + 0.7 \left |
      r_{\rho^+\pi^-} \right |^2 } \right ] 
\Bigg\} \, \%\, ,
\end{split}
\end{equation}
with 
\begin{equation}
r_{\rho^-\pi^+} = \frac{1 + \left (0.99 - 0.10 \, i \right ) \Delta
  C_{1/3}^{uu}}{1 + \left (0.96 - 0.12 \, i \right ) (\Delta
  C_{1/3}^{uu})^\ast} \,, \qquad
r_{\rho^+\pi^-} = \frac{1 + \left (0.96 + 0.07 \, i \right ) \Delta
  C_{1/3}^{uu}}{1 + \left (0.97 + 0.08 \, i \right ) (\Delta
  C_{1/3}^{uu})^\ast} \,.
\end{equation}
We have defined $\Delta C_{1/3}^{pp^\prime} \equiv \Delta C^{pp^\prime}_2+\Delta C^{pp^\prime}_1/3$
and the central values correspond to the default parameter choice employed in
\cite{Beneke:2003zv}, apart from $\beta$ for which we use $21.8^\circ$
\cite{Charles:2004jd}.  The result depends very strongly on the angle $\gamma$,
for which we use $\gamma =(70.0 \pm 10)^\circ$, in line with the current
world average \cite{Amhis:2012bh} obtained from $B^+\to DK^+$ and related
processes.  The quoted error is derived from the procedure used in
\cite{Beneke:2003zv} and is dominated by the dependence on $\gamma$.  We have
also included in the error the range of values obtained in scenarios S2 to S4
of that work.  The current experimental results are \cite{Amhis:2012bh}
\begin{align} 
S_{\pi^+  \pi^-} = -(65 \pm 6) \% \,,
\end{align} 
and \cite{Lees:2013nwa} 
\begin{align} 
S_{\rho\pi} = (5 \pm 9) \% \,. 
\end{align} 

The 90\%~CL constraints imposed by $S_{\pi^+ \pi^-}$ are displayed in brown in
the panels of Figure~\ref{fig:DC12uuG} and those by $S_{\rho \pi}$ in red.  For
the case of $\Delta C_2^{uu}$ one sees that combining the restrictions from
$R_{\pi^- \pi^0}$ with those from the two indirect CP asymmetries singles out
three allowed regions, one of which is the SM-like solution. The restriction on
$\Delta C_1^{uu}$ is much weaker than that on $\Delta C_2^{uu}$, which can be
understood evaluating (\ref{eq:C1C2ev}) numerically to find
\begin{align}
  \label{eq:C13}
  \Delta C_{1/3}^{uu}(m_b) &
  \simeq 1.0  \hspace{0.5mm} \Delta C_2^{uu}(M_W) 
       + 0.12 \hspace{0.5mm} \Delta C_1^{uu}(M_W) \,.
\end{align}
The quantity $\Delta C_1^{uu}(M_W)$ is multiplied by a small coefficient, so
that constraints from quantities such as these indirect CP asymmetries, which
only depend on this combination of coefficients, are roughly 10 times stronger
for $\Delta C_2^{uu}$ than for $\Delta C_1^{uu}$, when the method of varying
only one coefficient at a time is used. While the quantity $S_{\pi^+ \pi^-}$
still cuts out a very small portion of the allowed parameter space for $\Delta
C_1^{uu}$, it turns out that $S_{\rho\pi}$ offers no further constraint on
$\Delta C_1^{uu}(M_W)$ and has therefore been omitted from the figure.

Part of the remaining parameter space can be eliminated by
the quantity $R \left (\rho^- \rho^0/\rho^+ \rho^- \right )$, which is the ratio
of branching ratios of $B^-\to \rho^-\rho^0$ and $\bar B^0 \to \rho^+
\rho^-$.  The extension of the QCD factorisation formalism necessary to describe
these decays to NLO has been derived in~\cite{Beneke:2006hg, Bartsch:2008ps}.
The results for these decays depend on the parameters $\alpha_{1,2}(\rho\rho)$,
analogous to (\ref{eq:alpha1alpha2}) and as in the $\pi\pi$ sector these are
known to NNLO from \cite{Bell:2009fm, Beneke:2009ek}.  The decays also receive
contributions from QCD and electroweak penguin coefficients, which have only
been calculated up to the NLO level.  Combining all known corrections, one finds
\begin{eqnarray} 
  \label{eq:Rrho}
R \left (\rho^-
  \rho^0/\rho^+ \rho^- \right ) = \left ( 0.65^{+0.16}_{-0.11} \right
) \left [ \frac{1 + 2.1 \hspace{0.5mm} {\rm Re} \hspace{0.5mm} \Delta
    C_{+}^{uu} + 0.06 \hspace{0.5mm} {\rm Im} \hspace{0.5mm} \Delta
    C_{+}^{uu} + 1.1 \, \big | \Delta C_{+}^{uu} \big | ^2 }{1 + 2.0
    \hspace{0.5mm} {\rm Re} \hspace{0.5mm} \Delta C_{1/3}^{uu} + 0.23
    \hspace{0.5mm} {\rm Im} \hspace{0.5mm} \Delta C_{1/3}^{uu} + 1.0
    \, \big | \Delta C_{1/3}^{uu} \big | ^2 } \right ] . \hspace{8mm}
\end{eqnarray}
The corresponding experimental value is \cite{Amhis:2012bh}
\begin{align}
  R \left (\rho^- \rho^0/\rho^+ \rho^- \right ) & = 0.89 \pm 0.14 \,.  
\end{align}
The 90\% CL constraints from $R \left(\rho^-\rho^0/\rho^+ \rho^-\right)$ are shown in 
the green regions in Figure~\ref{fig:DC12uuG}.  
In the case of  $\Delta C_1^{uu}$ the constraint is only moderately useful, cutting out a small 
part of the otherwise allowed region.   The constraint on $\Delta C_2^{uu}$, on 
the other hand, serves to completely eliminate one of the regions where the phase
differs vastly from the SM value.\footnote{It is
worth mentioning that the central value in (\ref{eq:Rrho}) depends rather
strongly on the value of the hadronic input parameter $\lambda_B$: we use the
value of \cite{Bell:2009fm} quoted in Table \ref{tab:numeric:input}.  Using
lower values of $\lambda_B$ brings the SM number closer to the experimental
result \cite{Beneke:2006hg}.}

This concludes our discussion of constraints on the Wilson coefficients $\Delta
C_{1,2}^{uu}$. In Figure~\ref{fig:DC12uuG} the combined constraints on the
Wilson coefficients are overlaid on contours showing the ratio
$\Delta\Gamma_d/\Delta\Gamma_d^{\rm SM}$.  The allowed region for $\Delta C_{1}^{uu}$
contains areas where $\Delta\Gamma_d$ is enhanced by up to  30\% and also those
where it is decreased significantly compared to its SM value.  The allowed region for
$\Delta C_{2}^{uu}$ around the SM value does not allow for any significant changes 
in the ratio $\Delta\Gamma_d/\Delta\Gamma_d^{\rm SM}$, while that where the phase
is significantly different contains areas where the ratio is increased by
approximately 50\% compared to the SM value.   

%
%
\subsection{Bounds on up-charm-quark operators}
\label{sec:uc}

\begin{figure}[!t]
\begin{center}
\vspace{-5mm}
  \includegraphics[height=0.425 \textwidth]{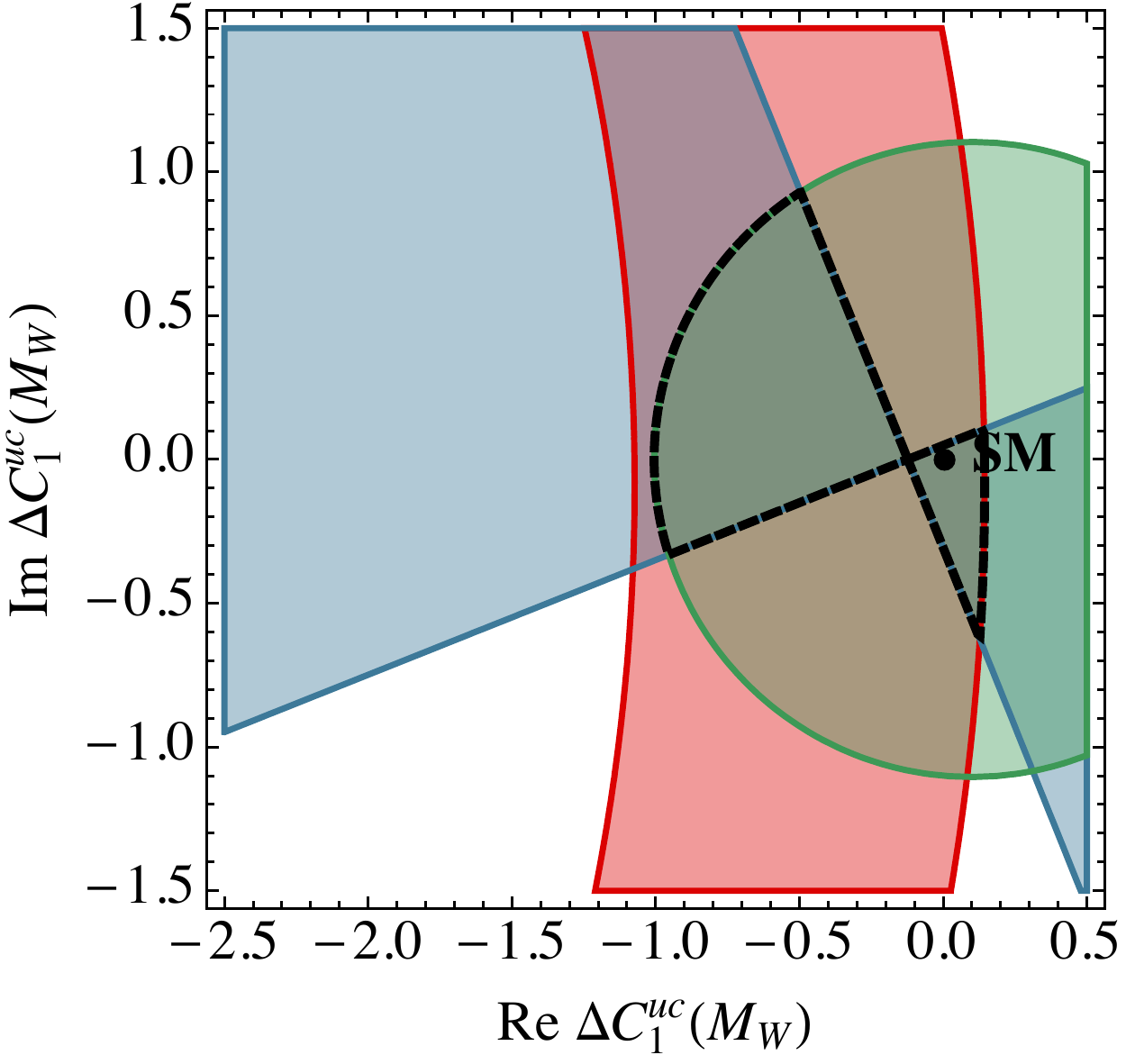} \qquad
  \includegraphics[height=0.425 \textwidth]{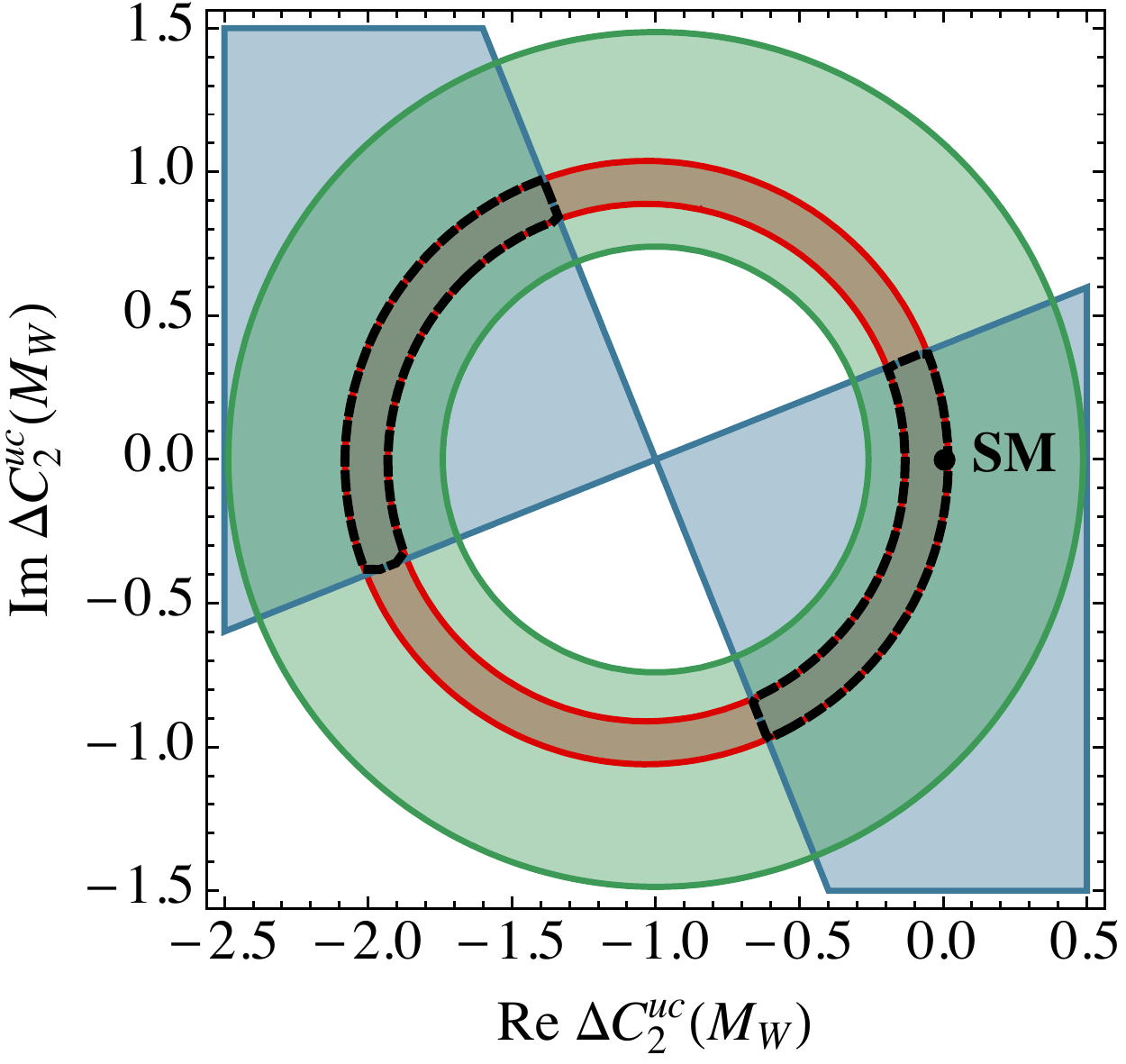} \\[4mm]
  \includegraphics[height=0.425 \textwidth]{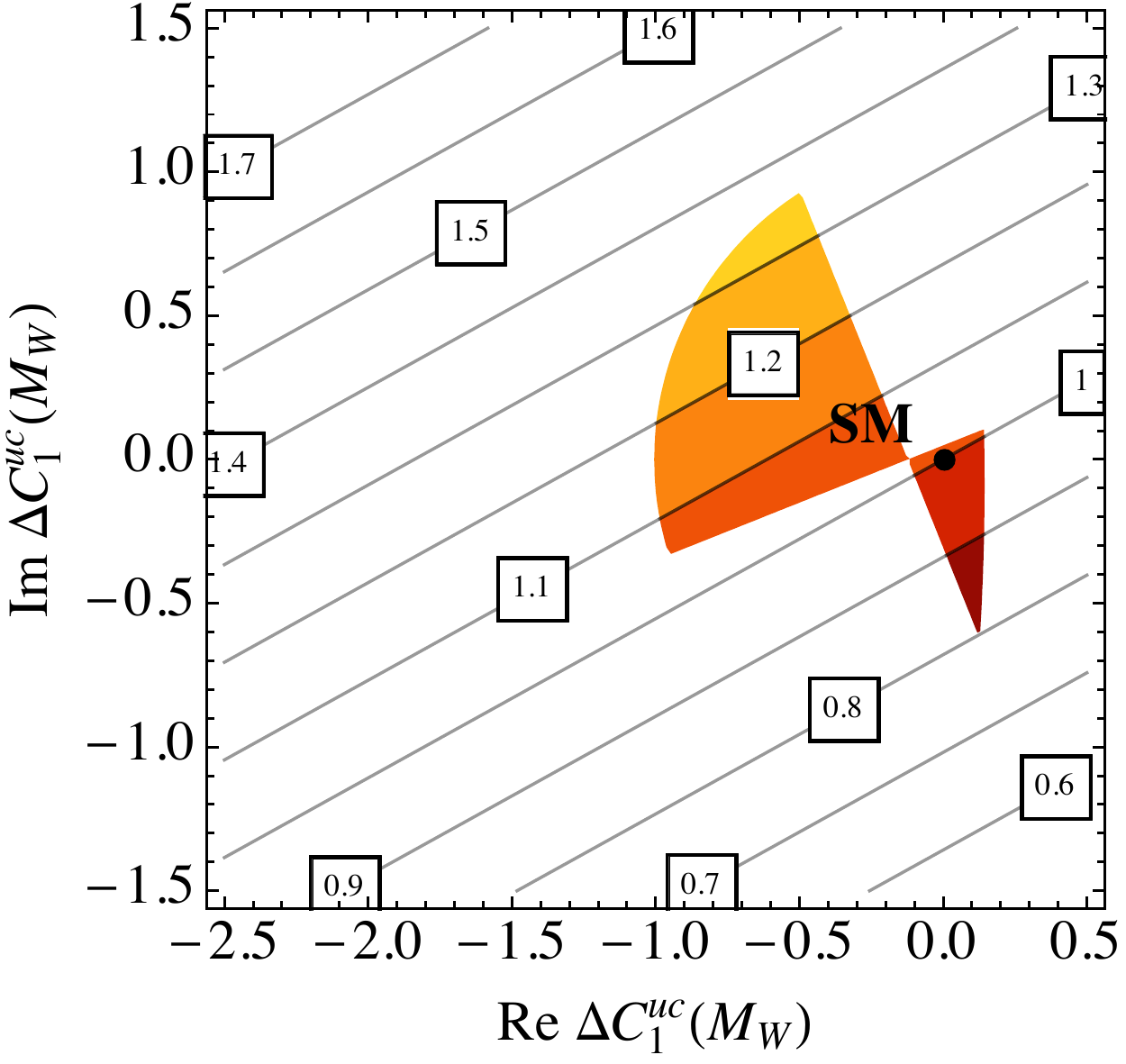} \qquad 
  \includegraphics[height=0.425 \textwidth]{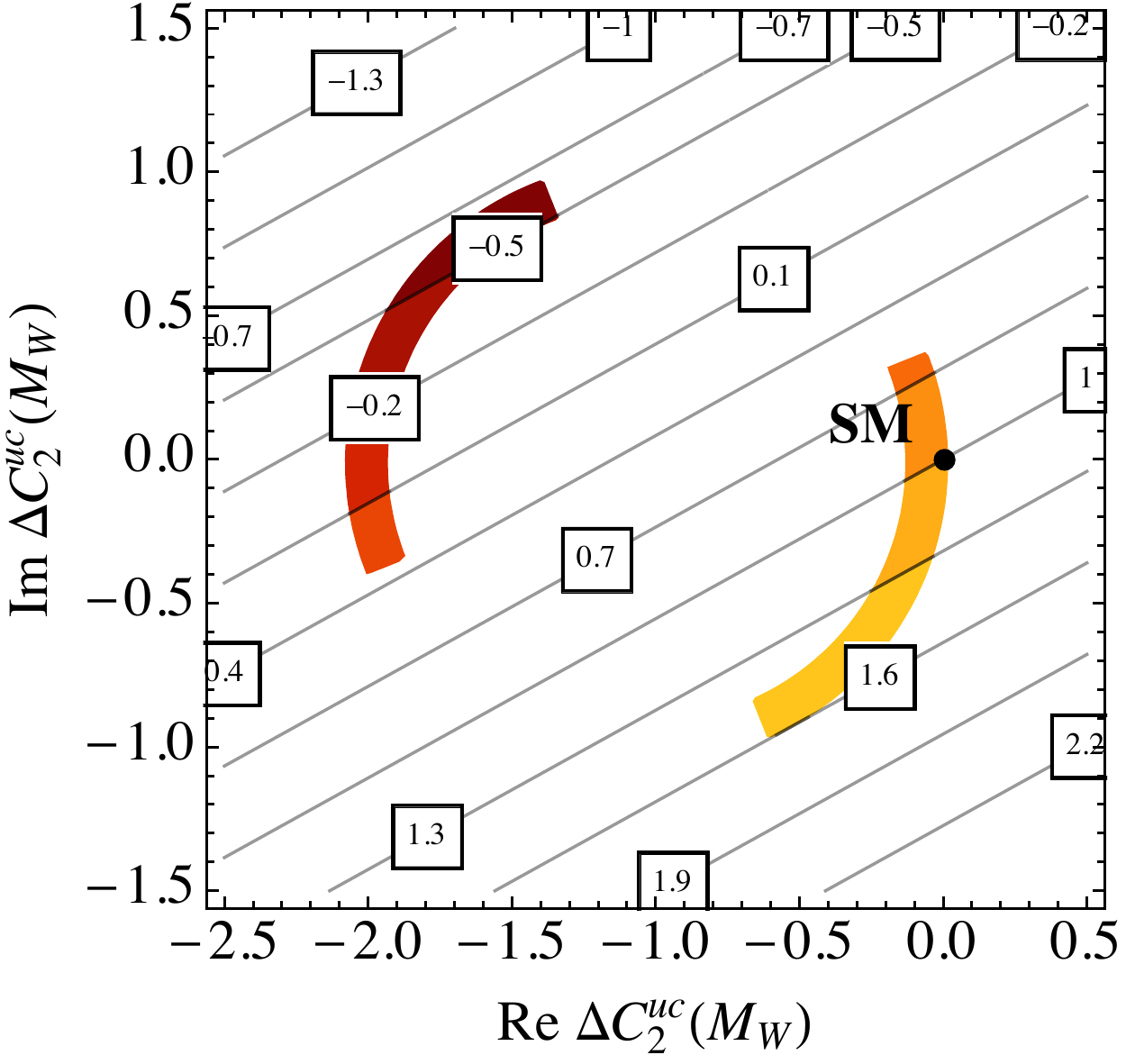}
  \caption{Upper panels: allowed parameter space in the ${\rm Re} \sp025 \Delta
    C_{1,2}^{uc}\sp025$--$\sp025{\rm Im} \sp025 \Delta C_{1,2}^{uc}$ planes. The
    red regions are related to constraints from $R_{D^{\ast +} \pi^-}$, the
    green ones from the total $B$-meson decay width and the blue ones from
    indirect CP asymmetries in exclusive $b\to c\bar{u} d$ decays. The regions
    enclosed by the dashed black lines represent the combined constraint from
    the different observables.  Lower panels: contours of
    $\Delta\Gamma_d^{}/\Delta\Gamma_d^{\rm SM}$ along with the combined
    constraints.  }
  \label{fig:DC12ucG}
\end{center}
\end{figure}

The Wilson coefficients $C_{1,2}^{uc}$ multiply the operators governing $b\to c
\bar{u} d$ transitions.  Such operators mediate exclusive decays such as $B\to
D\pi$ and are also important for the total $B$-meson decay rate, which receives
its largest contribution from $b\to c \bar{u} d$ decays.  In this section we
consider constraints from these two processes.

We first examine $B\to D\pi$ decays.  A particularly clean test of the
current-current sector can be obtained by relating the $\bar B^0 \to D^{\ast +}
\pi^-$ decay rate to the differential semi-leptonic $\bar B^0 \to D^{\ast +} l^-
\bar \nu_l$ rate normalised at $q^2 = m_{\pi^-}^2$. One has \cite{Beneke:2000ry}
\begin{align}
  \label{eq:RD} 
  R_{D^{\ast +} \pi^-} & 
  = \frac{\Gamma (\bar B^0 \to D^{\ast +} \pi^-)}
         {d\Gamma (\bar B^0 \to D^{\ast +} \ell^- \bar \nu_\ell)/dq^2 \big |_{q^2=m_{\pi^-}^2}} 
  \simeq 6 \pi^2 \sp025 f_{\pi}^2 \hspace{0.75mm} |V_{ud}|^2 \, \big | \alpha_2 (D^\ast \pi) \big |^2 \,, 
\end{align} 
with 
\begin{align} \alpha_2 (D^\ast \pi) = 1.054 + 0.013 \, i
+ \Delta C_{1/3}^{uc} \,.
\end{align} 
The SM value is obtained at NLO from \cite{Beneke:2000ry}, while the new-physics
contribution is LO accurate only. After restoring the theoretical error in the SM
calculation it follows that
\begin{eqnarray}
R_{D^{\ast +} \pi^-} \simeq (1.07 \pm 0.04 ) \, \Big [ 1+ 1.90
\hspace{0.5mm} {\rm Re} \sp025  \Delta C_{1/3}^{uc} + 0.02
\hspace{0.5mm} {\rm Im} \sp025  \Delta C_{1/3}^{uc} + 0.90
\hspace{0.5mm} \big | \Delta C_{1/3}^{uc} \big |^2 \Big ]  \, {\rm
  GeV}^2 \,. \hspace{6mm}
\end{eqnarray}
On the experimental side one has~\cite{Fleischer:2010ca}
\begin{align}
 R_{D^{\ast +} \pi^-} = (0.96
\pm 0.08 ) \, {\rm GeV}^2 \,.  
\end{align}
In Figure~\ref{fig:DC12ucG} we display in red the 90\% CL constraints on $\Delta
C_{1,2}^{uc}(M_W)$ obtained from $R_{D^{\ast +} \pi^-}$. Obviously, the allowed
region is the one which leaves the quantity $|\alpha_2(D^\ast \pi)|$ SM-like and is
thus a circle in the complex plane.  Moreover, since the result is sensitive
only to the combination $\Delta C_{1/3}^{uc}(m_b)$, the constraints on $\Delta
C_{1}^{uc}(M_W)$ are quite weak, which follows from~(\ref{eq:C13}).

In addition to the ratio of branching ratios above, we can also consider
indirect CP asymmetries in colour-suppressed $b\to c \bar{u}d$
transitions. These have been measured, for instance in \cite{Aubert:2007mn},
which gave results for $B^0\to D^{(\ast)0}h^0$ decays, where $h^0$ is a $\pi^0$,
$\eta$ or $\omega$ meson and the $D^{(\ast)0}$-meson decays into a CP eigenstate.  In
that case the amplitudes which determine the indirect asymmetry
through (\ref{eq:Sdef}) are dominated by $b\to c \bar{u} d$ transitions, because
the contribution from the $b\to u\bar{c} d$ is CKM suppressed.  In the SM, where
the Wilson coefficients $C^{uc}_{1,2}$ are real, the indirect asymmetry is
directly proportional to $\sin \left (2\beta \right)$ when the $b\to u\bar{c} d$
transitions are neglected. Beyond the SM the Wilson coefficients governing
the decay are in general complex and one must calculate the dependence of the
decay amplitudes on $C_{1,2}^{uc}$ in order to obtain the indirect asymmetry.
To the best of our knowledge there are no QCD factorisation predictions for such
decay amplitudes.  Resorting to an approximation in ``naive factorisation'',
however, we can estimate the indirect CP asymmetry to be 
\begin{align}
\label{eq:SDtheory}
S_{ D^{(\ast)0}h^0} = 2 \; \frac{{\rm Im} \left  (e^{-2i\beta }\rho_{ D^{(\ast)0}h^0} \right )}{1+|\rho_{ D^{(\ast)0}h^0}|^2} \,,
\qquad \rho_{ D^{(\ast)0}h^0} = \frac{C_1^{uc}+C_2^{uc}/3}{(C_1^{uc})^\ast + (C_2^{uc})^\ast /3} \,.
\end{align}
In this approximation the indirect asymmetry is independent of the decay channel
and can be compared with the experimental average from \cite{Aubert:2007mn}
\begin{align}
S_{ D^{(\ast)0}h^0} = -\left ( 56 \pm 24 \right)\% \,.
\end{align}
We wish to combine these two results into bounds on the Wilson coefficients, but
in order to do so we must assign an error to the theoretical estimate
(\ref{eq:SDtheory}). There is no obvious way to do this, so for lack of a better
procedure we assign a theory error equal to the experimental one above.  The
90\% CL bounds on $\Delta C_{1,2}^{uc}(M_W)$ obtained using these results are
shown in the blue regions of Figure~\ref{fig:DC12ucG}.  We will discuss the
utility of these constraints below.

In addition to exclusive decay modes, one can also examine constraints from
the total $B$-meson decay rate.  In fact, the largest contribution to the total
decay rate comes from the $b\to c \bar{u}d$ contribution, which probes $\Delta
C_{1,2}^{uc}$. Contributions from the $b\to c\bar{c}d$, $b\to u\bar{u}d$ and
$b\to u\bar{c} d$ modes are suppressed by at least a factor of 40 compared to
this dominant contribution and thus the total decay rate is largely insensitive
to changes in the Wilson coefficients associated with these contributions, i.e.~$\Delta C_{1,2}^{cc}$, $\Delta C_{1,2}^{uu}$ and $\Delta C_{1,2}^{cu}$
respectively.  Working to NLO in $\alpha_s$ and to $\Lambda_{\rm QCD}^2/m_b^2$
in the HQE, one finds 
(based on the numerical evaluation in \cite{Krinner:2013cja})
\begin{align}
\Gamma_{\rm tot}^{\rm SM} = (3.6\pm 0.8)\cdot 10^{-13} \, {\rm GeV}, \qquad 
\frac{\Gamma_{b\to c\bar{u}d}^{\rm SM}}{\Gamma_{\rm tot}^{\rm SM}}\simeq 44\% \,. 
\end{align}
We use these numbers to arrive at an approximate result for the total decay rate
as a function of the Wilson coefficients $C_{1,2}^{uc}$:
\begin{align}
  \frac{\Gamma_{\rm tot}}{\Gamma_{\rm tot}^{\rm SM}} & 
  \simeq
  \left(1-\frac{\Gamma^{\rm SM}_{b\to c\bar{u}d}}{\Gamma_{\rm tot}^{\rm SM}}\right)
  + \frac{3|C_1^{uc}|^2+3|C_2^{uc}|^2 + 2 \sp025 {\rm Re}\left[(C_1^{uc})^\ast C_2^{uc}\right]} 
         {\big(3|C_1^{uc}|^2+3|C_2^{uc}|^2 
          + 2 \sp025 {\rm Re}\left[(C_1^{uc})^\ast C_2^{uc}\right] \big)_{\rm SM}} 
  \, \frac{\Gamma_{b\to c\bar{u}d}^{\rm SM}}{\Gamma^{\rm SM}_{\rm tot}} \,.
\end{align}
The approximation is obtained by singling out the combination of Wilson
coefficients which multiplies the $b\to c \bar u d$ decay rate at LO in the HQE
and is thus valid to that accuracy. The experimental result is \cite{Amhis:2012bh}
\begin{align}
  \Gamma_{\rm tot} & = (4.20\pm 0.02) \cdot 10^{-13} \, {\rm GeV} \,.
\end{align}
The 90\% CL bounds on $\Delta C^{uc}_{1,2}(M_W)$ resulting from the above
equations for the total $B$-meson decay width are shown in green in
Figure~\ref{fig:DC12ucG}. It is evident that the total decay rate adds no
further constraint on the coefficient $\Delta C_2^{uc}$ compared to that from
the exclusive decays discussed above.  On the other hand, it is quite useful in
constraining $\Delta C_1^{uc}$, since it singles out the area of parameter space
where the phase is SM-like.

The lower panels of Figure~\ref{fig:DC12ucG} show the combined 90\% CL
constraints on $\Delta C_{1,2}^{uc}(M_W)$ derived above, overlaid on contours of
the ratio $\Delta\Gamma_d/\Delta\Gamma_d^{\rm SM}$.  In the allowed region for
$\Delta C_{1}^{uc}$, $\Delta\Gamma_d$ can be increased (decreased) by around
30\% (20\%) compared to its SM value.  In the allowed region for $\Delta C_{2}^{uc}$,
on the other hand, $\Delta\Gamma_d$ can be increased by up to 60\%, but can also
be decreased and even reach negative values.  These statements are independent
of whether constraints from the indirect CP asymmetries, which are on weaker
theoretical ground than the other two constraints, are included in the analysis.

%
%
\subsection{Bounds on charm-charm-quark operators}
\label{sec:cc}

The Wilson coefficients $C_{1,2}^{cc}$ multiply the operators governing $b\to
c\bar c d$ transitions.  These coefficients are especially important for
$\Delta\Gamma_d$, since they determine the contribution to~$\Gamma^d_{12}$
involving two charm quarks, which as seen from (\ref{eq:G12split}) dominates the
real part of this quantity.  However, the $b\to c\bar c d$ quark decay provides
only a negligible contribution to the total $B$-meson decay width and branching
ratios for decays into hadronic final states are under poor control
theoretically. This situation leads us to investigate indirect constraints from
$B\to X_d\gamma$ decays and the restrictions imposed by the dimension-eight
contributions to $\sin \left (2 \beta \right)$.  We also discuss indirect CP
asymmetries in exclusive $b\to c\bar{c}d$ transitions and the semi-leptonic
asymmetry $a_{sl}^d$.
  
\begin{figure}[!t]
\begin{center}
\vspace{-5mm}
  \includegraphics[height=0.42 \textwidth]{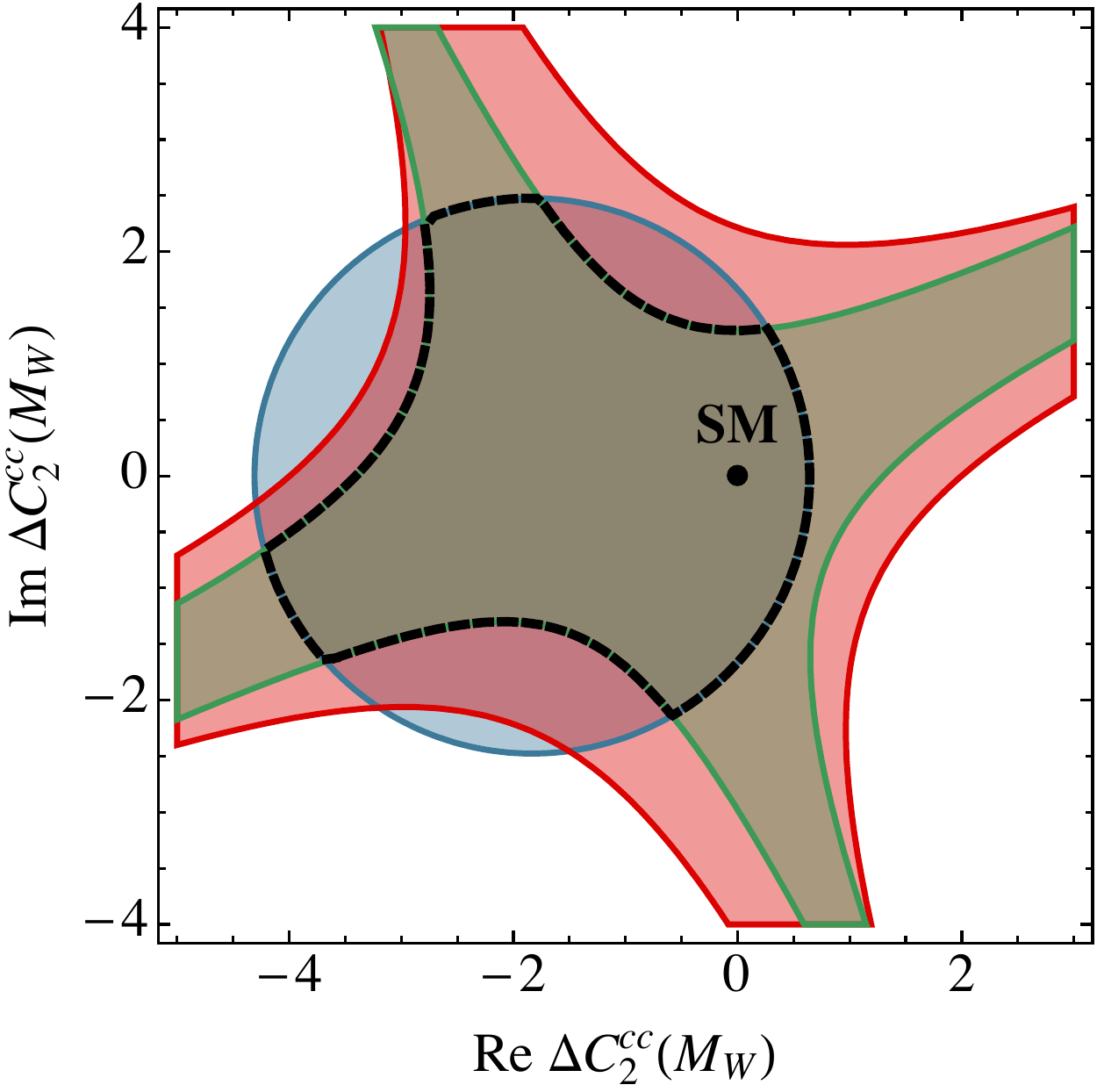} \qquad \quad 
  \includegraphics[height=0.42 \textwidth]{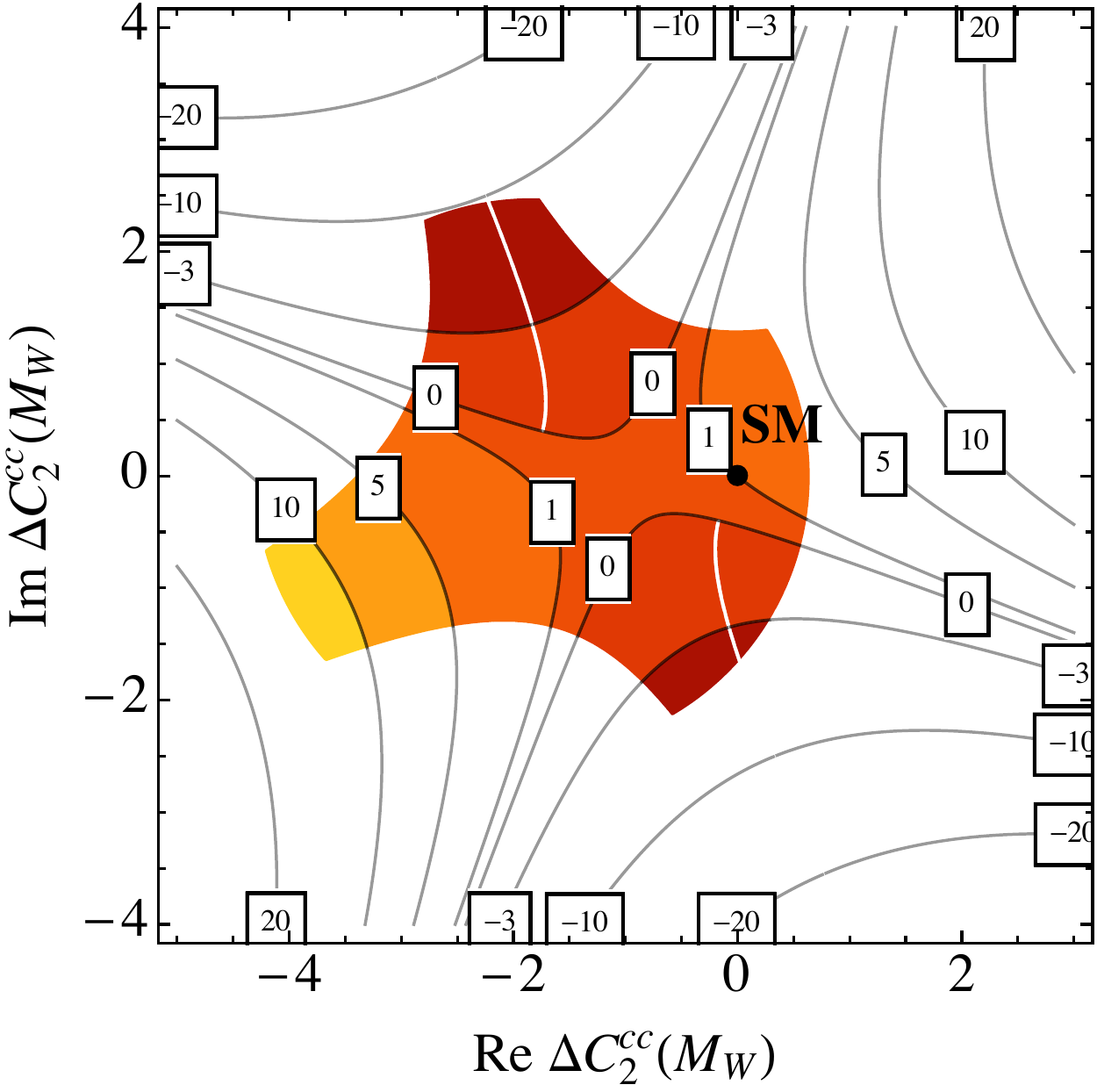}
  \caption{Left panel: allowed parameter space in the ${\rm Re} \sp025 
    \Delta C_{2}^{cc}\sp025 $--$\sp025 {\rm Im} \sp025 
    \Delta C_{2}^{cc}$ planes. The blue circular region is related to constraints
    from $B\to X_d \gamma$,  and the green region from those from 
    semi-leptonic asymmetry $a_{sl}^d$, and the red region from the dimension-eight
    contributions to $\sin \left (2\beta \right )$.   The region
    enclosed by the dashed black lines represent the combined constraint from
    the different observables.  
    Right panel: contours of $\Delta\Gamma_d^{}/\Delta\Gamma_d^{\rm SM}$.
  }
  \label{fig:DC12ccG} 
\end{center}
\end{figure}

We begin by considering $B\to X_d\gamma$.  The branching ratio for this decay
is sensitive to the Wilson coefficients $C_{1,2}^{cc}(M_W)$ even at leading
logarithmic (LL) order in
perturbation theory, due to operator mixing. To clarify this point, we follow
\cite{Gambino:2001ew} and write
\begin{align}
  \label{eq:P0def}
  {\rm Br} \left ( B \to X_d\gamma \right )_{E_\gamma > E_0} &
  = {\cal N}_{B \to X_d\gamma} \hspace{0.5mm} \left|\frac{\lambda_t^d}{V_{cb}}
    \right|^2 \, \big [ P(E_0)+ N(E_0) \big ] \,,
\end{align}
where ${\cal N} _{B \to X_d\gamma} = 2.57\cdot10^{-3}$ is a normalisation
factor, $P(E_0)$ is proportional to the perturbative expression for the partonic
decay $b\to X_d\gamma$, $N(E_0)$ denotes power-suppressed nonperturbative
contributions and $E_0$ repesents a cut on the photon energy $E_\gamma$. At LL accuracy the
perturbative contribution is given by $P(E_0)=\big |C^{(0)}_7(m_b) \big |^2$,
where $C_7$ is the Wilson coefficient of the electromagnetic dipole operator
(see (\ref{eq:Q7Q9}) below). In terms of matching coefficients at the scale
$M_W$, one finds
\begin{align}
  C_7^{(0)}(m_b) \simeq -0.31 - 0.17 \hspace{0.5mm} \Delta C_2^{cc}(M_W) \, .
\end{align}
The following approximation for the branching ratio is thus valid 
\bea
\label{eq:LLconstraint}
\begin{split}
  {\rm Br} \left (B \to X_d\gamma \right )_{E_\gamma > E_0} &
  \simeq {\rm Br} \left (B \to X_d\gamma \right )_{E_\gamma > E_0}^{\rm SM} 
\\[2mm]
& \phantom{xx}
  + \frac{{\cal N}_{B \to X_d\gamma} }{10} \left|\frac{\lambda_t^d}{V_{cb}}\right|^2 
 \Big( 1.1 \,{\rm Re}\, \Delta C_2^{cc}(M_W) 
       + 0.3 \, |\Delta C_2^{cc}(M_W)|^2 
 \Big) \,. \hspace{6mm}
\end{split}
\eea
We can obtain bounds on the Wilson coefficient $\Delta C^{cc}_2(M_W)$ using the
above equations, together with the recent SM prediction for the CP-averaged
branching ratio \cite{Crivellin:2011ba}
\begin{align}
  \label{eq:BdgSM}
  {\rm Br} \left (B \to X_d \gamma \right )^{\rm SM}_{E_\gamma>1.6 \,{\rm GeV}} &
  = \left (1.54^{\small +0.26}_{\small -0.31} \right )\cdot 10^{-5} \,,
\end{align}
and the latest experimental result
\cite{delAmoSanchez:2010ae,Crivellin:2011ba} 
\begin{align}
  \label{eq:BdgExp}
  {\rm Br} \left (B \to X_d \gamma \right )_{E_\gamma>1.6 \,{\rm GeV}} &
  = (1.41 \pm 0.57)\cdot 10^{-5} \,.
\end{align}
The 90\% CL bound on $\Delta C_2^{cc}(M_W)$ resulting from this
analysis is shown by the blue circle in Figure~\ref{fig:DC12ccG}.
Obviously, the LL expression (\ref{eq:LLconstraint}) does not
translate into severe constraints.  It could be slightly improved by
noting that even though the correction to the $B\to X_d \gamma$ decay
rate from the matrix element of $Q_2^{cc}$ first appears at NLO in
$\alpha_s$, it is nonetheless sizable due to the fact that the SM
coefficient $C_2^{\rm SM} (m_b)\simeq 1.1$ is about 4 times larger
than $C_7^{\rm SM}(m_b) \simeq -0.31$ in magnitude.  Moreover, the
virtual corrections from this operator are more important than those
related to real emission. We have examined how the constraints shown
in the figure change when adding on these NLO virtual corrections to
(\ref{eq:LLconstraint}).  It turns out that the allowed region is
moderately reduced, but not to the degree that it would change the
qualitative analysis, so we do not discuss modifications of
(\ref{eq:LLconstraint}) due to higher-order QCD corrections any
further.

The results from $B\to X_d \gamma$ branching ratio are of little
use for constraining potential new-physics contributions to
$\Delta\Gamma_d$, as is verified by comparing the allowed region in
the blue circle with the contours of
$\Delta\Gamma_d/\Delta\Gamma_d^{\rm SM}$ in the right-hand panel of
Figure~\ref{fig:DC12ccG}.  We have checked that even weaker
constraints arise from $A_{\rm CP} (B\to X_{d+s}\gamma)$, $B^+\to
\pi^+ \mu^+\mu^-$ and the $B$-meson lifetime.  In principle, we could
also consider indirect asymmetries in exclusive $b\to c\bar{c}d$
decays, analogous to those for the $b\to c \bar{u}d$ transitions
discussed above.  However, in the present case the difficulties in
estimating theory uncertainties related to hadronic matrix elements
are compounded by the presence of penguin amplitudes carrying a
different weak phase to the current-current contributions.  We have
checked that even being relatively aggressive with theory errors
yields a constraint broadly similar to the blue triangular regions in
Figure~\ref{fig:DC12ucG}, which in any case does not help in ruling
out areas of parameter space where $\Delta\Gamma_d$ is dramatically
different from the SM value.  For these reasons, we do not use these
indirect asymmetries in our analysis.

The situation described above leads us to seek
additional experimental constraints on the $\Delta C_{i}^{cc}$ coefficients from
observables in the $B_d$-meson mixing system itself.  One such constraint arises
from the precise determination of $\sin \left ( 2 \beta \right )$ (see
(\ref{eq:NPphaseint})) from the time-dependent CP asymmetry in $B \to J/\psi
K_S$ decays.  Within the SM the latter observable is to excellent precision simply
given by the imaginary part of the dimension-six contribution to $M_{12}^d$ that
arises from box diagrams with top-quark and $W$-boson exchange.  Additional
dimension-eight contributions  stem  from double insertions of $\Delta B =1$ operators 
such as the current-currents operators. 
Including both types of corrections, but restricting ourselves for the moment to insertions of
$Q_{1,2}^{cc}$, we obtain
\bea
  \label{eq:M12d}
\begin{split}
  M_{12}^d & 
  \simeq  \frac{G_F^2}{12\sp025 \pi^2} \hspace{0.5mm} M_{B_d} \hspace{0.5mm} f_{B_d}^2\hspace{0.5mm} B_V \hspace{1mm} \Bigg \{  ( \lambda_t^d \big )^2
  m_t^2 \, K_6  
\\[2mm] 
  & \hspace{0.5cm}  
  + ( \lambda_c^d \big )^2 \, m_b^2 \, \bigg [ \big ( C_2^{cc} \big)^2  K_8^{cc}
 + \left (  2 \sp025  C_2^{cc} \sp025   C_1^{cc}  
   + 3 \sp025  \big ( C_1^{cc} \big)^2  \right  )  \widetilde{K}_8^{cc} \bigg ]
  \ln \left ( \frac{M_W^2}{m_b^2}  \right )  \! \Bigg \} \,.
\end{split}
\eea
Here and in the remainder of the section all Wilson coefficients are to be evaluated at the  
scale $\mu = M_W$. After employing  $C_{2}^{cc} \simeq 1 + \Delta C_2^{cc}$ and 
$C_{1}^{cc} \simeq \Delta C_1^{cc}$ we confirm the results of~\cite{Boos:2004xp}
for the SM contribution. The coefficient that multiplies the dimension-six
contribution is given at next-to-leading logarithmic~(NLL) accuracy by $K_6
\simeq 0.47$, while we left the leading logarithm of the dimension-eight
contribution unresummed, which is sufficient for our purposes. The corresponding
dimension-eight coefficients take the form 
\begin{align}
  \label{eq:K8}
  K_8^{cc} & 
  = \frac{1}{3} - \frac{2 \sp025 m_c^2}{m_b^2} + \frac{5 \hspace{0.5mm} B_S}{12 \sp025  B_V}
  \simeq 0.82 \,, &
  \widetilde{K}_8^{cc} & 
  = \frac{1}{3} - \frac{2 \sp025  m_c^2}{m_b^2} - \frac{\widetilde{B}_S}{12 \sp025  B_V} 
  \simeq -1.1 \cdot 10^{-3} \,, 
\end{align}
where we have used the central values as given  in Table \ref{tab:numeric:input}
of the so-called bag parameters to obtain the final numerical values. Varying
$B_V$, $B_S$ and $\widetilde B_S$ within their uncertainties we find that 
$|\tilde K_8^{cc}/K_8^{cc}| < 0.05$, which implies that the contribution to 
$M_{12}^d$ involving $C_1^{cc}$ is generically suppressed by more than an
order of magnitude. The dimension-eight contribution involving only up quarks is obtained
from (\ref{eq:M12d}) and (\ref{eq:K8}) by replacing $ ( \lambda_c^d \big )^2 \to
( \lambda_u^d \big )^2$, $C_{1,2}^{cc} \to C_{1,2}^{uu}$ and $m_c^2 \to 0$,
while in the case of the charm-up-quark contribution one has to employ $
( \lambda_c^d \big )^2 \to 2 \lambda_c^d \lambda_u^d$, $\big(C_{2}^{cc}\big)^2 
\to C_{2}^{cu} \sp025 C_{2}^{uc}$, $2 \sp025 C_2^{cc} \sp025 C_1^{cc} + 3 \sp025 
\big (C_1^{cc}\big)^2 \to C_2^{cu} \sp025 C_1^{uc} + C_2^{uc} \sp025 C_1^{cu} + 
3 \sp025 C_1^{cu} \sp025 C_1^{uc}$ and $m_c^2 \to m_c^2/2$. Notice that since
$m_c^2/m_b^2 \simeq 0.1$ and $K_8^{pp^\prime} \gg \widetilde K_8^{pp^\prime}$
the coefficients $C_2^{pp^\prime}$ are for each flavour combination much
stronger constrained by the measurement of the time-dependent CP asymmetry in $B
\to J/\psi K_S$ than~$C_1^{pp^\prime}$.

The correction $\phi_d^\Delta$ to the $B_d$-mixing phase $2\beta$ is related to
the off-diagonal element $M_{12}^d$ of the mass matrix via
\begin{align}
  \label{eq:S2BNP}
  \sin \left ( 2 \beta + \phi_d^\Delta \right ) &
  = {\rm Im} \left ( \frac{M_{12}^d}{|M_{12}^d|} \right ) \,.
\end{align}
We can obtain a simple analytic expression for the new-physics phase by approximating
$\lambda_t^d \simeq |\lambda_t^d| \hspace{0.5mm} e^{i\beta}$,
$\lambda_c^d \simeq - |\lambda_c^d|$ and $|\lambda_c^d/\lambda_t^d| \simeq1$,
expanding in powers of $m_b^2/m_t^2$, ignoring contributions proportional to
$\widetilde K_8^{cc}$ and neglecting the tiny SM contribution to $\phi_d^\Delta$
of ${\cal O} (10^{-4})$~\cite{Boos:2004xp}. We then find from (\ref{eq:M12d}) 
\begin{align}
  \label{eq:S2Bapprox}
  \phi_d^\Delta & 
  \simeq -\frac{m_b^2}{m_t^2} \sp025 \frac{K_8^{cc}}{K_6}  
  \Big[ \sin \left ( 2 \beta \right )  {\rm Re} \left ( C_2^{cc} \right )
      - \cos \left ( 2 \beta \right )  {\rm Im} \left ( C_2^{cc} \right ) \Big]
  \ln \left ( \frac{M_W^2}{m_b^2}  \right ) .  \hspace{4mm}
\end{align}
By using in addition $\lambda_u^d = - \lambda_t^d - \lambda_c^d$ similar results
can be shown to hold in the case of the charm-up-quark and up-up-quark
contribution.  

Current experimental and SM results for the CKM angle $\beta$ can be found in the last row of Table~\ref{tab:mixing}. We would like to use these to constrain new-physics contributions to $\sin \left ( 2 \beta \right)$, but run into the problem that the experimental and SM results are in poor agreement. One way to deal with this would be to bridge the gap through the new-physics part of the dimension-eight contributions above. However, it is not clear that this discrepancy merits a new-physics explanation. To derive 90\% CL constraints from this observable, we instead use the following procedure. First, we ignore the difference in central values of the SM and experimental numbers, and add the experimental and theory errors together in quadrature to derive a total error. We then saturate the total error at 90\% CL with the new-physics contribution from (\ref{eq:S2BNP}). This procedure leads to the constraint on $\Delta C_2^{cc}$ shown by the red region in the left panel in Figure~\ref{fig:DC12ccG}, which we have obtained using the exact numerical evaluation of (\ref{eq:M12d}) and (\ref{eq:S2BNP}) rather than the approximation (\ref{eq:S2Bapprox}). As mentioned above, the constraint on $\Delta C_1^{cc}$ is very weak, as are those on the coefficients $\Delta C_{1,2}^{uu}$ and $\Delta C_{1,2}^{uc}$, so we omit these from the discussion. We see that the constraint derived from the dimension-eight contribution to $\sin \left ( 2 \beta \right)$ does very little to eliminate the allowed parameter space. In fact, this is an important result in its own right --- if these corrections had turned out to be restrictive, there would be little justification for ignoring dimension-eight operators in other parts of our analysis.

We can derive a more useful restriction by studying the semi-leptonic asymmetry $a^d_{sl}$. The new-physics contributions to this quantity can be calculated in a straightforward way using
the parameterisation (\ref{eq:GammaNP}), including the new-physics contributions
related to $M_{12}^d$ in~(\ref{eq:M12d}) as well as those from (\ref{eq:dg12}).
The recent experimental results from direct measurements of semi-leptonic decays
have been given in Table~\ref{tab:mixing}.  The numbers are consistent with zero
within the large experimental errors, so for simplicity we use as an
experimental value $\left(0\pm 1\right)\%$ in deriving bounds on the Wilson
coefficients.  The allowed region resulting from this procedure is shown in blue
in Figure~\ref{fig:DC12ccG}.  While some of the allowed space from $B\to
X_d\gamma$ is cut out upon including this constraint, ${\cal O}(10)$
contributions $\Delta\Gamma_d$ due to the Wilson coefficient $\Delta C_2^{cc}$
are not ruled out from present experimental data, as is shown in the right-hand panel 
of the figure. 

%
%
\section{New physics in $\bm{\Delta \Gamma_d}$: 
         $\bm{\left(\bar{d} b\right) \left(\bar{\tau} \tau\right)}$ operators}
\label{sec:DGNPtautau}

In this section we study possible effects on $\Delta\Gamma_d$ related to
effective operators of the form $\left(\bar{d}
  b\right)\left(\bar{\tau} \tau\right)$. The analogous operators for
the $B_s$-meson system (i.e.~those obtained by replacing $d \to s$) were
introduced in \cite{Bobeth:2011st} and used in studying
$\Delta\Gamma_s$. 
The corresponding effective Hamiltonian reads 
\begin{align}
  \label{eq:Leffsbtautau} 
  {\cal H}_{\rm eff}^{b\to d \tau^+ \tau^-} & 
  = - \frac{4G_F}{\sqrt{2}} \, \lambda_t^d \, \sum_{i,j} C_{i,j} (\mu) \Op_{i,j}
  \, ,  
\end{align}
and involves the following complete set of operators
\begin{equation}
\label{eq:tauops}
\begin{split}
  Q_{S, AB} & = ( \bar d \, P_A \, b )  (\bar \tau \, 
              P_B \, \tau  ) \,, 
\\[1mm]
  Q_{V, AB}  & =  ( \bar d \, \gamma^\mu P_A \, b  )  (\bar \tau \, 
                 \gamma_\mu    P_B \, \tau  ) \,, 
\\[1mm]
  Q_{T, A }  & =  ( \bar d   \, \sigma^{\mu\nu}  P_A \, b  )    (\bar \tau\, 
                  \sigma_{\mu\nu}   P_B \, \tau  ) \,, 
\end{split}
\end{equation}
where $P_{L,R}=(1\mp \gamma_5)/2$ and $A,B=L,R$. In addition to these operators,
our analysis will use the dimension-six effective Hamiltonian describing $b\to d
\ell^+\ell^-$ transitions ($\ell = e,\, \mu,\, \tau$).  We write this as
\begin{align}
  {\cal H}^{b \to d \ell^+ \ell^-}_{\rm eff} & 
  = - \frac{4G_F}{\sqrt{2}}  \, \lambda_t^d \, \sum_i C_i(\mu) Q_i
  \, .
\end{align}
The most important operators in what follows are
\begin{equation}
  \label{eq:Q7Q9}
\begin{split}
 Q_7 & = \frac{e}{(4\pi)^2} \hspace{0.5mm} m_b \left(\bar d \, \sigma^{\mu\nu}P_R \, b\right) \, F_{\mu\nu}  \,, \\[1mm]
 Q_{9} &= \frac{e^2}{(4\pi)^2} \left(\bar d \, \gamma^{\mu}P_L \, b\right) 
                             \left(\bar \ell \, \gamma_\mu \, \ell \right) \,, \\[1mm]
 Q_{10} & = \frac{e^2}{(4\pi)^2} \left(\bar d \, \gamma^{\mu}P_L \, b\right)
                                 \left( \bar \ell \,  \gamma_\mu\gamma_5 \, \ell \right)\,,
\end{split}
\end{equation}
and their chirality-flipped counterparts $Q_i^\prime$ obtained through the
interchange $P_L\leftrightarrow P_R$. 

The operators (\ref{eq:tauops}) are interesting because they can give large
contributions to $\Delta\Gamma_{d}$, but are only weakly constrained by
experimental data.  We will see that in the case of the $B_d$-meson system the
various direct and indirect bounds on the Wilson coefficients of the operators
(\ref{eq:tauops}) are generally weaker than in the $B_s$-meson system and that
large enhancements of $\Delta\Gamma_d$ due to such operators are not yet ruled
out.  We derive direct bounds, where the operators contribute to tree-level
matrix elements, from the decays $B_d \to \tau^+\tau^-$, $B \to X_d
\tau^+\tau^-$ and $B^+ \to \pi^+ \tau^+ \tau^-$. Indirect bounds, where the
operators contribute either through operator mixing and loop-level matrix
elements, are based on $B \to X_d\gamma$ and $B^+ \to \pi^+ \mu^+\mu^-$
decays. We discuss the two cases in turn.

%
%
\subsection{Direct bounds}
\label{sec:direct}

We first investigate direct constraints from the decay $B_d \to \tau^+\tau^-$.
At present, the only experimental information on this decay is the 90\% CL bound
from \cite{Aubert:2005qw}:
\begin{equation}
  \label{eq:Btt_Bound}
  {\rm Br} \left (B_d \to \tau^+ \tau^- \right ) < 4.1 \, \cdot \, 10^{-3} \,.
\end{equation}
The theory prediction, including both SM and the effects of the operators
(\ref{eq:tauops}) can be extracted from \cite{Bobeth:2002ch}.  The result
depends on the SM coefficient $C_{10}$ in addition to the Wilson coefficients
$C_{S,AB}$ and $C_{V,AB}$, but not on the tensor coefficients
$C_{T,A}$. Moreover, due to a loop suppression factor of $\alpha/\pi$, the SM
contribution alone is quite small, $\mbox{Br} \left (B_d \to \tau^+ \tau^-
\right )_{\rm SM} \simeq 2.3 \cdot 10^{-8}$ \cite{Bobeth:2013uxa}. We therefore
neglect it in obtaining bounds on the coefficients of the new
operators. Furthermore, as before we assume the dominance of a single operator
at a time, neglecting interference terms of the new operators both with the SM
contribution and with themselves.  Following this procedure, we can set bounds
on the absolute values of the coefficients $C_{S,{AB}}$ and $C_{V,{AB}}$,
independent of the chirality structure.  Giving up this assumption, would lead
to considerably larger bounds on $\Delta \Gamma_d$. In that respect our
estimates are very conservative.

The branching ratio depends on a number of input parameters.  Some of these are
common to the other decays discussed in this section: for these we use the
values of $G_F,\, M_{B_d},\, \tau_{B_d},\, f_{B_d},\, |\lambda_d^t|,\, m_\tau,\,
\alpha(M_Z),\, m_b^{\rm pole}$ and $m_d$ summarised in
Table~\ref{tab:numeric:input}. We then obtain at $m_b = \overline m_b (\overline
m_b) \simeq 4.2$ GeV the results $|C_{S,AB}(m_b)| < 1.1$ and
$|C_{V,AB}(m_b)|<2.2$, which are also collected in
Table~\ref{tab:direct_bounds}.  These are the strongest bounds on the scalar and
vector coefficients that will emerge from our analysis.

We next consider inclusive and exclusive $b\to d \tau^+ \tau^-$ decays.  In this
case, there are no direct experimental bounds on the branching ratios.  However,
we can use information from the $B_{d,s}$-meson lifetimes to estimate the
potential size of new-physics contributions to these decays.  We first note that
the SM prediction for the lifetime ratio is very close to unity
\cite{Lenz:2011ti}
\begin{equation}
  \left( \frac{\tau_{B_s}}{\tau_{B_d}} -1\right)_{\rm SM} = 
  \left ( -0.2 \pm 0.2 \right ) \% \, .
\end{equation}
Current experimental measurements \cite{Amhis:2012bh} are compatible with this
prediction:
\begin{equation}
  \left( \frac{\tau_{B_s}}{\tau_{B_d}} -1 \right)
  = \left ( -0.2 \pm 0.9 \right ) \% \, . 
\end{equation}
Comparing the results, one can get a rough bound on the size of
possible new-physics contributions $\Gamma_{d,s}^{\rm NP}$ to the total
$B_{d,s}$-meson  decay rates $\Gamma_{d,s}$, namely
\begin{equation}
  \frac{\Gamma_d^{\rm NP} - \Gamma_s^{\rm NP}}{\Gamma_s} 
  = \left ( 0.0 \pm 0.9 \right ) \% \; .
\end{equation}
Setting $\Gamma_s^{\rm NP}=0$ gives an upper bound on (also invisible)
new-physics contributions to $B_d$-meson decays. At $90 \% \, {\rm CL}$ one
obtains
\begin{equation}
  \label{eq:estimates}
       {\rm Br}\left  (B_d \to X \right ) \leq 1.5\% \,.
\end{equation}  

We now turn this estimate into bounds on Wilson
coefficients using the theoretical expressions for the $B \to X_d \tau^+\tau^-$
and $B^+ \to \pi^+ \tau^+\tau^-$ branching ratios.  In contrast to $B_d\to
\tau^+\tau^-$ decays, in these cases all operators contribute, so we also gain
information on the tensor coefficients.  However, once again the results are
independent of the chirality structure and allow us to constrain only absolute
values of the coefficients. For the inclusive decay $B \to X_d \tau^+\tau^-$, we
use the expressions for the branching ratios given in Section~3 and the appendix
of \cite{Bobeth:2011st} after appropriate modifications.  Most of the inputs to
the branching ratio are common to $B_d$-meson and $B_s$-meson decays.  Apart
from trivial differences related to meson masses and lifetimes (for the
exclusive decay we use $\tau_{B^+}$ and the CKM factor $|V_{td}^\ast
V_{tb}^{}|/|V_{cb}^{}|$ given in Table \ref{tab:numeric:input}). The exclusive
decay $B^+ \to \pi^+ \tau^+\tau^-$ depends on these parameters and also three
$B\to \pi$ form factors ($f_{+,T,0}$), as a function of the dilepton invariant
mass, denoted as $q^2$.  For these we use the results of
\cite{Khodjamirian:2011ub}.\footnote{We have also derived bounds using the form
  factors from \cite{Ball:2004ye}, which yields similar but slightly more
  stringent bounds.}  Moreover, we integrate over the full kinematic range $q^2
\in [4\sp025 m_{\tau}^2, (M_{B+}-M_{\pi^+})^2]$.

We must also decide on a value for the experimental branching ratios.  At
present, we can only use (\ref{eq:estimates}) to estimate the size of the $B \to
X_d \tau^+\tau^-$ and $B^+ \to \pi^+ \tau^+\tau^-$ branching ratios, which most
likely overestimates their allowed ranges. Here a direct experimental bound
would be very helpful.  For reference, we collect the obtained bounds on the
Wilson coefficients using the $90\% \, {\rm
  CL}$ estimates in Table~\ref{tab:direct_bounds}.  In Section~\ref{sec:DGtau},
we will show the size of possible enhancement of $\Delta\Gamma_d$ as a function
of measured branching ratios.  Compared to the bounds on the $B\to X_d
\tau^+\tau^-$ and $B^+\to \pi^+\tau^+\tau^-$ branching ratios estimated through
(\ref{eq:estimates}), we find tiny SM predictions
\beq
\begin{split}
  \mbox{Br} \left (B\to X_d \tau^+\tau^- \right )_{\rm SM} & 
  = (1.2 \pm 0.3) \cdot 10^{-8} \,,
\\[2mm]
  \mbox{Br}\left (B^+\to \pi^+ \tau^+\tau^- \right )_{\rm SM} & 
  = (1.5 \pm 0.5) \cdot 10^{-8} \,.
\end{split}
\eeq
In both cases our results refer to the fully-integrated and non-resonant
branching ratios. The inclusive decay includes just the LO corrections, but
accounts for contributions proportional to the tau mass that are of kinematic
origin \cite{Guetta:1997fw}.  As for the exclusive mode, we stayed within naive
factorisation and the error reflects the uncertainty due to the use of different
$B\to \pi$ form-factor determinations \cite{Khodjamirian:2011ub,Ball:2004ye}.

%
%
\subsection{Indirect bounds}

Indirect bounds arise from cases where the operators (\ref{eq:tauops}) do not
give tree-level contributions to the decays, but contribute either through
operator mixing, through loop-level matrix elements or through both. The
theoretical expressions needed to set various indirect bounds can be adapted
from \cite{Bobeth:2011st}. It turns out that the most stringent indirect bounds
on the Wilson coefficients can be derived from $B \to X_d \gamma$ and $B^{+} \to
\pi^+ \mu^+\mu^-$ decays.  We have also examined constraints from $B_d\to\gamma
\gamma$ decays, but these are rather weak compared to the other decays and in
some cases they depend very strongly on hadronic input parameters, so we do not
discuss them further.

%
%

We first derive bounds from $B\to X_d\gamma$ decays.  The structure of branching
ratios for these decays has been discussed in Section~\ref{sec:cc}.  We find
that the quantity $P(E_0)$ in (\ref{eq:P0def}) can be written as
\begin{equation}
\begin{aligned}
  \label{eq:P0exp}
  P(E_0) &
   = \left|C_7^{{\rm SM}(0)}(m_b) + \Delta C^{\rm eff}_{7}(m_b)\right|^2 
   + \left|\Delta {C^{\rm eff}_{7}}^\prime(m_b)\right|^2 \,,
\end{aligned}
\end{equation}
where 
\begin{equation}
\begin{aligned} 
  \label{eq:C7C9eff}
  \Delta C_{7}^{{\rm eff}(\prime)} (m_b) & 
   = \sqrt{x_\tau}\left( 0.62 - \, 1.09 \, \eta_6^{-1}+4 \ln x_\tau\right) C_{T,R(L)} (m_b)
  \,. 
\end{aligned}
\end{equation}
We have defined $\sqrt{x_{\tau}} \equiv m_\tau/m_b^{\rm pole}$ and the quantity
$\eta_6 \equiv \alpha_s(\Lambda_{\rm NP})/\alpha_s(m_t)$, where $\alpha_s$ is to
be evaluated with six active flavours.  The first two terms in
(\ref{eq:C7C9eff}) arise from operator mixing and the third is the matrix
element of $Q_{T,A}$.  In order to derive 90\% CL constraints on the Wilson
coefficients $C_{T,A}(m_b)$, we insert (\ref{eq:P0exp}) into~(\ref{eq:P0def})
and compare with the experimental result (\ref{eq:BdgExp}), as usual considering
the dominance of one operator at a time. This procedure yields for $\Lambda_{\rm
  NP} \simeq 1$ TeV the bounds $|C_{T,R}(m_b)|< 0.2$ and $|C_{T,L}(m_b)|<0.1$,
which translate to $|\Delta C_{7}^{\rm eff}(m_b)| < 0.7$ and $|\Delta C_{7}^{\rm
  eff \prime}(m_b)| < 0.3$.

%
%

The rare decay $B^+\to \pi^+ \mu^+\mu^-$ has been observed by LHCb
\cite{LHCb:2012de} in the 2011 data sample of $ 1 \, {\rm fb}^{-1}$ with
$(25.3^{+6.7}_{-6.4})$ events. This provides the first measurement of the
non-resonant branching ratio
\begin{align}
  \label{eq:Btopimumu:exp}
  {\rm Br} \left (B^+\to \pi^+ \mu^+\mu^- \right )&
    = (2.3 \pm 0.6 ) \cdot 10^{-8} \,.
\end{align}
integrated over the whole dilepton invariant mass range $q^2 \in 
[4\sp025  m_\mu^2, (M_{B^+} - M_{\pi^+})^2]$.

In principle, the calculation of exclusive $B\to M \ell^+\ell^-$ ($M = P, V$)
decays is advanced, see \cite{Ali:2013zfa} for the most recent prediction of
$B^+\to \pi^+ \mu^+\mu^-$ in the SM, where $B\to \pi \ell \nu_\ell$ data has
been used to extract information on the form factors. The inclusion of
corrections beyond naive factorisation in QCD factorisation at low $q^2$ has
been discussed in \cite{Beneke:2004dp} (see also \cite{Dimou:2012un}), whereas at high $q^2$ a local operator
product expansion can be applied to account for resonant contributions 
\cite{Grinstein:2004vb, Beylich:2011aq}. However, in the absence of experimental
measurements for either region separately and in view of the large experimental
uncertainty, we evaluate the branching ratio in the naive factorisation
approximation following \cite{Bobeth:2001sq}. Employing the $B\to \pi$ form
factors of \cite{Khodjamirian:2011ub}, we obtain
\beq \label{eq:SMpred}
  {\rm Br} \left (B^+\to \pi^+\mu^+\mu^- \right )_{\rm SM}  = 
     (2.1 \pm 0.4) \cdot 10^{-8} \,.
\eeq
The given uncertainty encodes the error related to differences in the existing
$B\to \pi$ form-factor determinations~\cite{Khodjamirian:2011ub, Ball:2004ye}. 
Our prediction (\ref{eq:SMpred}) is close both to the measured 
value~(\ref{eq:Btopimumu:exp}) and the SM value obtained in \cite{Ali:2013zfa}.

Like in the case of the  inclusive decay $B\to X_d\gamma$, the effective operators
(\ref{eq:tauops}) contribute to $B^+\to \pi^+ \mu^+\mu^-$ via mixing into the
operators mediating $b\to d \gamma$ and $b\to d \ell^+\ell^-$~$(\ell =
e, \mu)$. The case of $b\to s$ transitions has been previously discussed in
\cite{Bobeth:2011st} and can be adopted with appropriate replacements to $b\to
d$ transitions.  One then finds contributions from the tensor coefficients
$C_{T,A}$, and also on the linear combination $(C_{V,AL} + C_{V,AR})$ of the
vector coefficients.  The scalar Wilson coefficients $C_{S,AB}$ are not subject
to constraints from $B^+\to \pi^+ \mu^+\mu^-$.

\begin{figure}[t!]
\centering
  \includegraphics[height=0.45\textwidth]{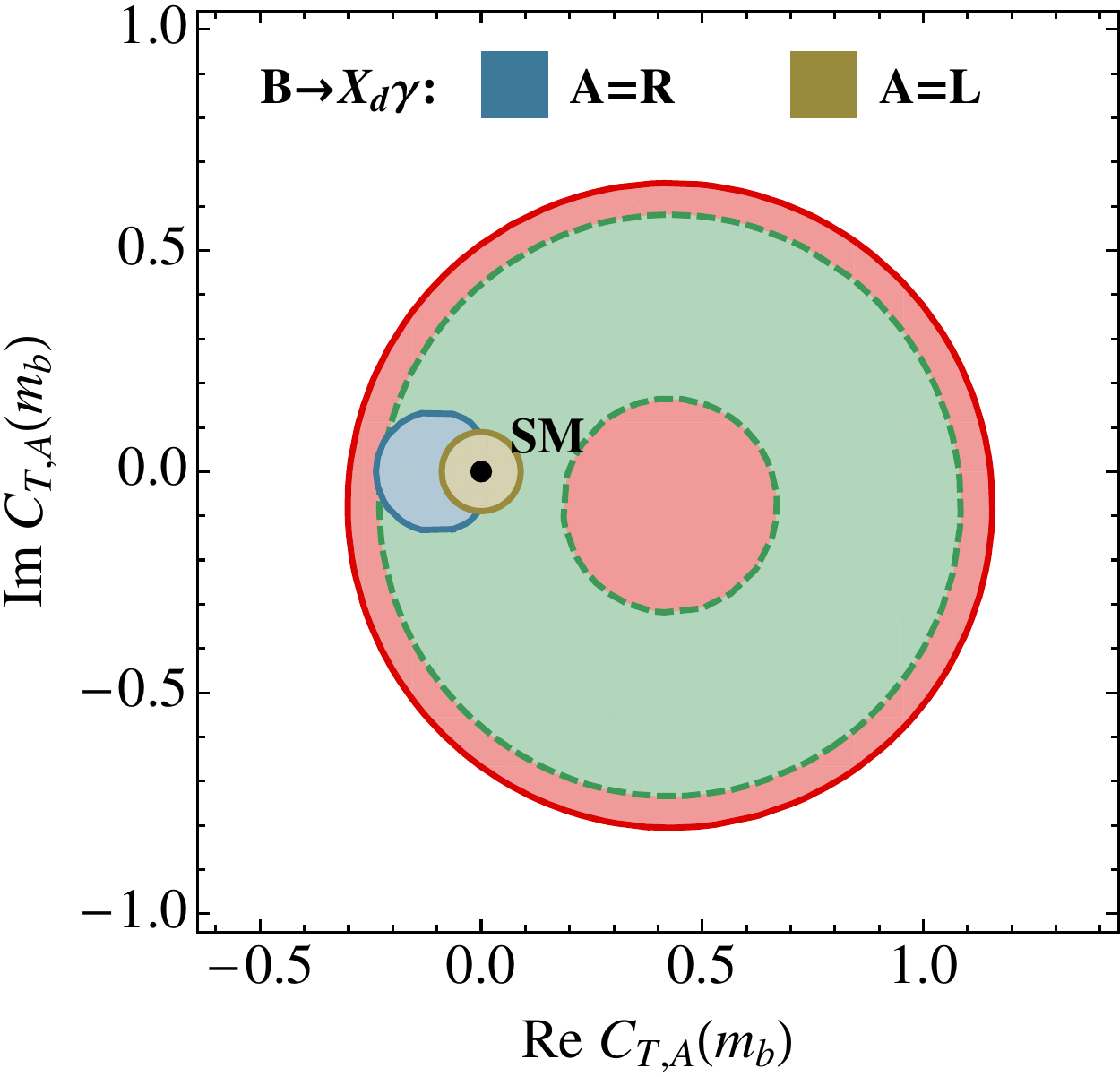}
  \hskip 0.05\textwidth 
  \includegraphics[height=0.45\textwidth]{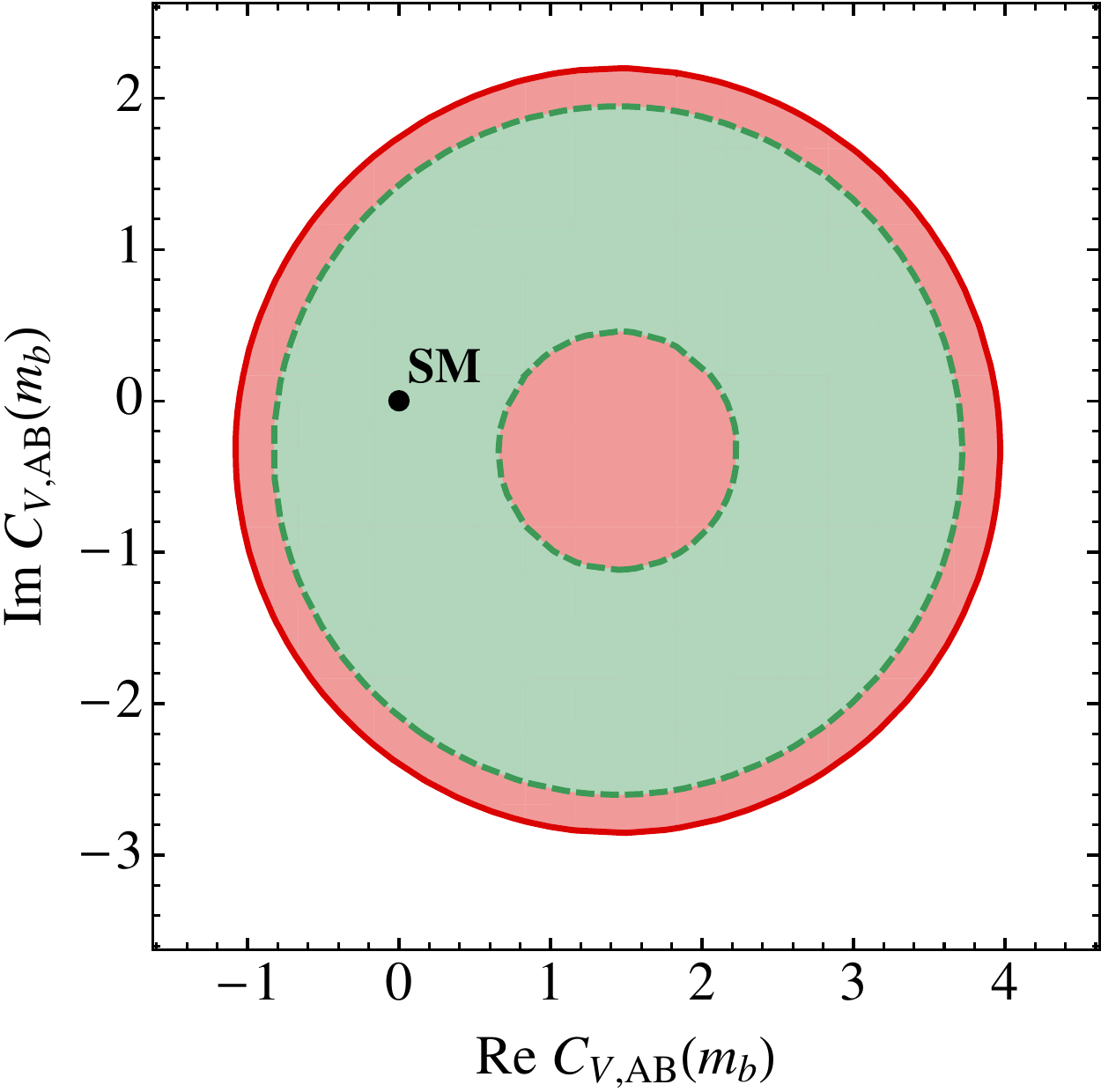}
  \caption{The 90\% CL regions of $C_{T,A}(m_b)$ (left) and $\big (C_{V,AL}(m_b)
      + C_{V,AR}(m_b) \big)$ (right) from ${\rm Br} \left (B^+\to
      \pi^+\mu^+\mu^- \right )$ (red) and ${\rm Br}\left (B\to X_d\gamma \right
    )$ for $T,A = T,R$ (blue) and $T,A = T,L$ (brown).  The prospects assuming a
    measurement of ${\rm Br} \left (B^+\to \pi^+\mu^+\mu^- \right )$ with $7$
    fb$^{-1}$ at LHCb are shown as dashed~(green) contours.  }
  \label{fig:Btopimumu}
\end{figure}

In Figure~\ref{fig:Btopimumu}, we show the 90\% CL regions in the planes of
complex-valued $C_{T,A}(m_b)$ and $\big ( C_{V,AL} (m_b) + C_{V,AR} (m_b) \big
)$ allowed by ${\rm Br} \left (B^+\to \pi^+\mu^+\mu^- \right)$. In the plots a
theory uncertainty of 20\% of the SM prediction is assumed and the form-factor
predictions \cite{Khodjamirian:2011ub} are used. We see that in the case of the
tensor Wilson coefficients, $B^+\to \pi^+\mu^+\mu^-$ provides at present the
constraint $|C_{T,A}(m_b)| \lesssim 1.2$, which as indicated is clearly weaker
than the sensitivity of the inclusive decay $B \to X_d \gamma$. Assuming single
operator dominance, the current constraint on the vector Wilson coefficients
$|C_{V,AB}(m_b)| \lesssim 4.0$ is not as strong as the one from $B_d \to
\tau^+\tau^-$.  For comparison we also
show contours assuming that LHCb has collected $7 \, {\rm fb}^{-1}$ of data by
2017. For this purpose the current statistical errors have been rescaled by a
factor $1/\sqrt{7}$. This exercise shows the potential of this decay mode to
reduce further the allowed ranges of $b\to d \tau^+\tau^-$ Wilson
coefficients. Depending on the central value of the measurement, it will provide
complementary constraints to $B\to X_d\gamma$ for the tensor Wilson
coefficients.

\begin{table}[t!]
\begin{center}
\renewcommand{\arraystretch}{1.1}
\begin{tabular}{|c|c|c|c|}
\hline
  Constraint & $|C_{S,AB}(m_b)|$ & $|C_{V,AB}(m_b)|$ & $|C_{T,A}(m_b)|$
\\
\hline
\multicolumn{4}{|c|}{direct}
\\
\hline
  ${\rm Br} \left (B_d\rightarrow \tau^+\tau^- \right )$        &  $1.1$   &  $2.2$  &  --- 
\\
  ${\rm Br} \left ( B\rightarrow X_d \tau^+\tau^- \right )$      &  $10.6$  &  $5.3$  &  $1.5$
\\
  ${\rm Br} \left ( B^+\rightarrow \pi^+ \tau^+\tau^- \right )$  &  $5.9$   &  $6.2$  &  $2.9$
\\
\hline 
\multicolumn{4}{|c|}{indirect}
\\
\hline
  \multirow{2}{*}{${\rm Br} \left ( B\rightarrow X_d \gamma \right )$}
& \multirow{2}{*}{---}
& \multirow{2}{*}{---} 
& $0.2$ for $A=R$
\\
   & & & $0.1$ for $A=L$
\\
  ${\rm Br} \left ( B^+\rightarrow \pi^+ \mu^+\mu^-\right)$  &  ---    &  $4.0$  &  $1.2$
\\
\hline
\end{tabular}
\renewcommand{\arraystretch}{1.0}
\caption{Summary of direct and indirect bounds on the Wilson coefficients
  (\ref{eq:Leffsbtautau}) at the bottom-quark mass scale $m_b = \overline{m}_b
  (\overline m_b) \simeq 4.2$ GeV.  The constraint from $B_d \to \tau^+\tau^-$
  decay follows from the experimental 90\%~CL bound ${\rm Br} \left (B_d \to
    \tau^+\tau^- \right ) < 4.1\cdot 10^{-3}$, whereas those from $B \to
  X_d\tau^+\tau^-$ and $B^+ \to \pi^+\tau^+\tau^-$ refer to the 90\%~CL estimate
  from (\ref{eq:estimates}). Note that the bounds are independent of the chiral
  structure $A,B=L,R$ unless explicitly indicated.  }
  \label{tab:direct_bounds}
\end{center}
\end{table}

%
%
\subsection{Maximal effects in width difference}
\label{sec:DGtau}

We now explore the consequences of the bounds on the Wilson coefficients
(\ref{eq:Leffsbtautau}) obtained in the previous section on the size of possible
enhancements in $\Delta\Gamma_d$. To do so we consider the parameter $|
\tilde{\Delta}_d|$ introduced in (\ref{eq:GammaNP}).  For the $B_s$-meson case
expressions for this quantity as a function of the relevant Wilson coefficients
were presented in \cite{Bobeth:2011st}, and we can make use of these results
after a trivial substitution of CKM factors. Assuming single operator dominance,
we then find
\begin{equation}
  \label{eq:DGNP}
\begin{aligned}
  |\tilde{\Delta}_d|_{S,AB} & < 1+ \left(0.41^{+0.13}_{-0.08}\right)\, |C_{S,AB}(m_b)|^2 \,,
\\[2mm]
  |\tilde{\Delta}_d|_{V,AB} & < 1+ \left(0.42^{+0.13}_{-0.08}\right)\, |C_{V,AB}(m_b)|^2 \,,
\\[2mm]
  |\tilde{\Delta}_d|_{T,A }\;\; & < 1+ \left(3.81^{+1.21}_{-0.74}\right)\, |C_{T,A}(m_b)|^2 \,,
\end{aligned}
\end{equation}
where the quoted uncertainties are related to the theory error of
$\Delta\Gamma_d^{\rm SM}$. The numerical input values of the bag parameters
$B_V$, $B_S$ and $\widetilde{B}_S$ are given in Table \ref{tab:numeric:input}.
Using the strongest bounds from Table~\ref{tab:direct_bounds},
i.e.~$|C_{S,AB}(m_b)| \lesssim 1.1$, $|C_{V,AB}(m_b)| \lesssim 2.2$,
$|C_{T,L}(m_b) |\lesssim 0.1$ and $|C_{T,R}(m_b)| \lesssim 0.2$, results in
\begin{align}
  |\tilde{\Delta}_d|_{S,AB} & \lesssim 1.6 \,, &
  |\tilde{\Delta}_d|_{V,AB} & \lesssim 3.7 \,, &
  |\tilde{\Delta}_d|_{T,L} & \lesssim 1.05 \,, &
  |\tilde{\Delta}_d|_{T,R} & \lesssim 1.2 \, \; .
\end{align}
These numbers imply that the scalar operators can lead to an enhancement of
about $60\%$ over the SM prediction, whereas in the case of vector operators
even deviations in excess of $270\%$ are allowed. The
possible deviations due to tensor operators can, on the other hand, amount to at
most $20\%$. Such small effects are undetectable given that  the hadronic uncertainty
in $\Delta \Gamma_d$ is of similar size.

\begin{figure}[t!]
\centering
  \includegraphics[width=0.48\textwidth]{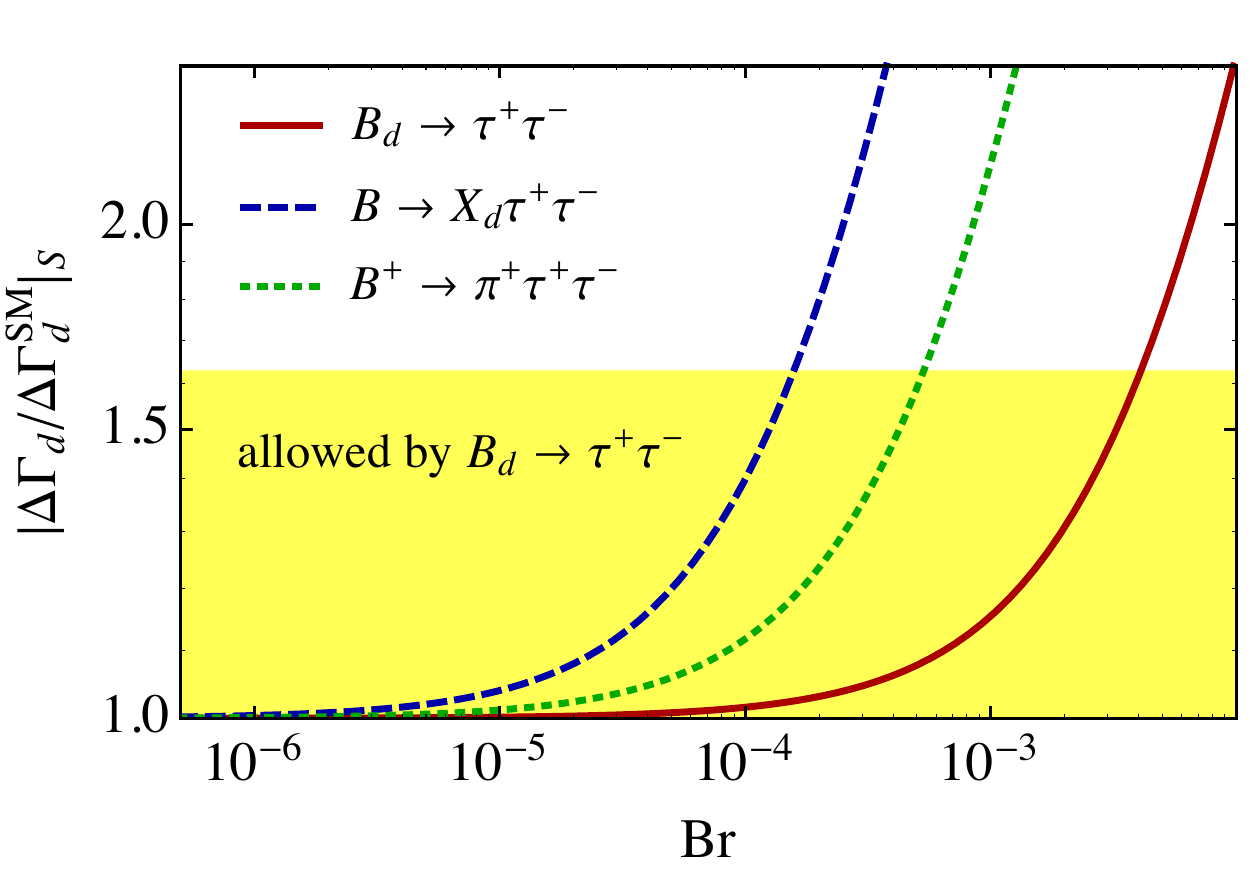}  \quad
  \includegraphics[width=0.48\textwidth]{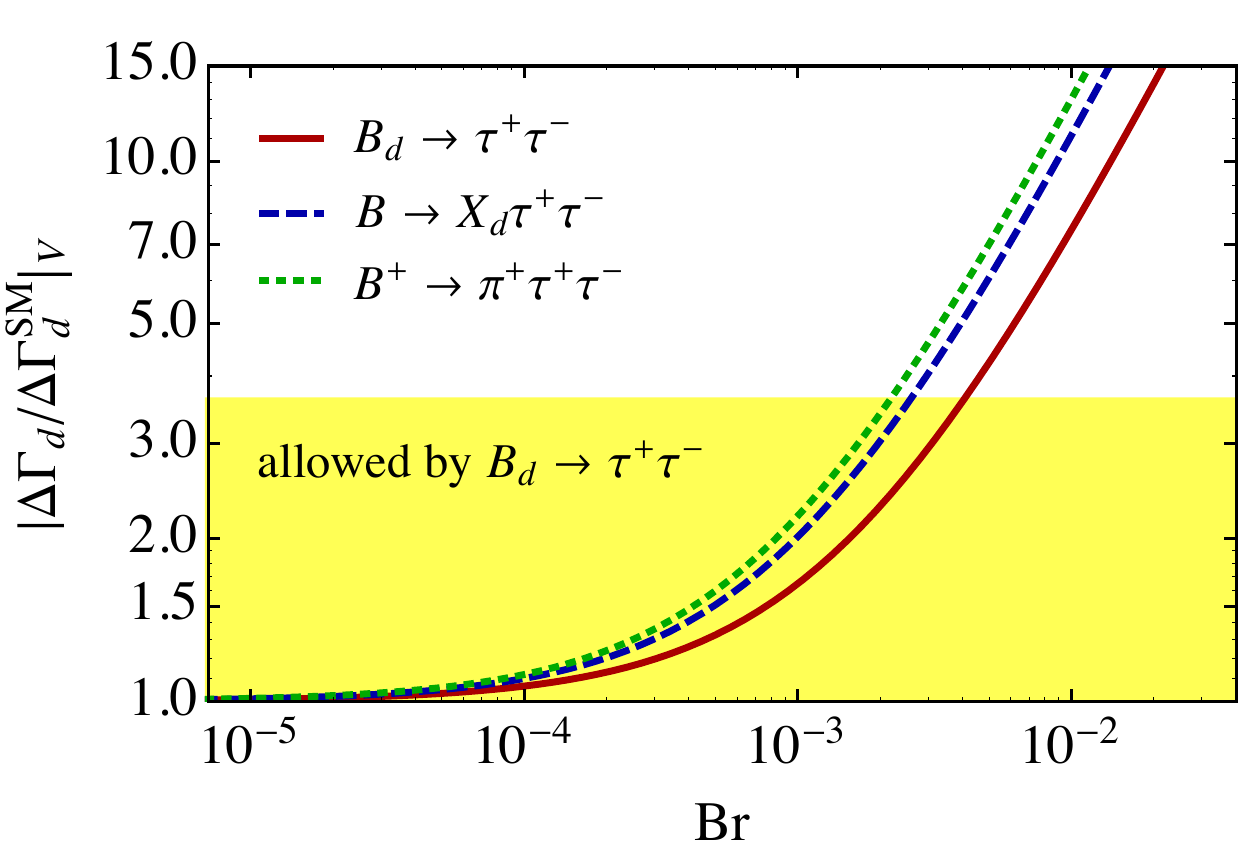} 
  \caption{90\% CL bounds on possible enhancements of $\Delta \Gamma_d$ induced
    by the different $(\bar{d}b) (\bar\tau\tau)$ operators.  The left panel
    shows the effect of scalar operators, while the right panel illustrates the
    case of vector operators.  In both panels the yellow region indicates the
    maximal enhancements that are consistent with (\ref{eq:Btt_Bound}).  The
    effect of an experimental improvement in the $B_d\to \tau^+\tau^-$, $ B \to
    X_d \tau^+ \tau^-$ and $B^+ \to \pi^+ \tau^+ \tau^-$ branching ratios is
    indicated by the solid red, the dashed blue and the dotted green curves,
    respectively.  }
  \label{fig:Br}
\end{figure}

It is also interesting to study the impact future improved extractions of ${\rm
  Br} \left (B_d \to \tau^+ \tau^- \right )$, ${\rm Br} \left ( B \to
  X_d\tau^+\tau^- \right )$ and ${\rm Br} \left ( B^+\to \pi^+ \tau^+\tau^-
\right )$ will have on the maximal enhancements in $\Delta \Gamma_d$.  Such a
comparison is provided in Figure~\ref{fig:Br} for the scalar operators (left
panel) and the vector operators (right panel). The plots show that for both the
scalar and the vector operators and fixed branching ratio the $B_d \to \tau^+
\tau^-$ decay always provides the most stringent constraint on $|\Delta
\Gamma_d/\Delta \Gamma_d^{\rm SM}|$. This implies that in order to restrict
possible new-physics effects in $\Delta \Gamma_d$, future measurements of the $B
\to X_d\tau^+\tau^-$ or $B^+\to \pi^+ \tau^+\tau^-$ branching ratio have to
surpass the present bound (\ref{eq:Btt_Bound}) on ${\rm Br} \left (B_d \to
  \tau^+ \tau^- \right )$. Numerically, we find that limits of ${\rm Br} \left (
  B^+ \to \pi^+ \tau^+ \tau^- \right ) \lesssim 5.3 \cdot 10^{-4}$ and ${\rm Br}
\left ( B \to X_d \tau^+ \tau^- \right ) \lesssim 2.6 \cdot 10^{-3}$ would be
required in the case of the scalar and vector operators to reach the current
sensitivity of the $B_d \to \tau^+ \tau^- $ branching ratio. 

%
%
\section{Conclusion}
\label{sec:conclusion}

In this article we have investigated the room for new-physics effects in the
decay rate difference $\Delta \Gamma_d$ of neutral $B_d$ mesons following an
effective field theory approach. Such a study is well-motivated because the
current direct experimental bound on $\Delta \Gamma_d$ still allows for an
enhancement of several 100\% over the SM prediction and a new measurement at
D\O{} can be interpreted as a solution of the longstanding problem with the
dimuon asymmetry, involving an anomalous enhancement of the decay
rate difference.

We have presented a detailed comparison between $\Delta \Gamma_d$ and $\Delta
\Gamma_s$ within the SM, emphasising that while in the former case the relevant
CKM factors $V_{cd}^\ast V_{cb}^{}$ and $V_{ud}^\ast V_{ub}^{}$ both scale as
$\lambda^3$, in the latter case there is a hierarchy between the individual
contributions, since $V_{cs}^\ast V_{cb} = {\cal O} (\lambda^2)$ and
$V_{us}^\ast V_{ub}= {\cal O} (\lambda^4)$. In consequence, a modification in $b
\to c \bar{c} d$ can have a much larger effect in $\Delta \Gamma_d$, compared to
the effect of a similar modification in $b \to c \bar{c} s$ on $\Delta
\Gamma_s$. Such modifications could for instance arise in new-physics scenarios
that predict violations of the CKM unitarity. If these non-standard corrections
affect the $B_d$-meson and $B_s$-meson sectors in a flavour-universal way, then
$\Delta \Gamma_d$ can be enhanced relative to the SM by up to $300\%$, while in
the case of $\Delta \Gamma_s$ the corresponding shifts can be $50\%$ at
most. Our general arguments show that for $\Delta \Gamma_d$ a large enhancement
is a priori not excluded, while in the case of $\Delta \Gamma_s$ beyond the SM
effects cannot dramatically exceed the size of the hadronic uncertainties that
plague the decay rate differences.

Our general findings have been corroborated by model-independent studies of
non-standard effects associated to dimension-six operators of flavour content
$(\bar{d} p)(\bar p^{\hspace{0.25mm} \prime} b)$ with $p, p^\prime = u,c$ as
well as $(\bar{d}b)(\bar\tau\tau)$. Allowing for flavour-dependent and complex
Wilson coefficients, we performed detailed analyses of the experimental
constraints on each individual coefficient of the current-current operators that
arise from hadronic $B$-meson decays like $B \to \pi \pi,\, \rho\pi,\,
\rho\rho,\, D^\ast \pi$, the inclusive $B\to X_d\gamma$ transition and the
dimension-eight contributions to the mixing-induced CP asymmetry in $B \to
J/\psi K_S$.  In accordance with our general observations, we found that large
relative deviations in $\Delta\Gamma_d$ and $a_{sl}^d$ of a few $100\%$ can at
present not be ruled out phenomenologically, if they are associated to the
current-current operator $(\bar{d} c)(\bar c b)$. We stressed that the derived
bounds also apply to new-physics models that feature violations of the unitarity
of the CKM matrix, a scenario more restrictive than our own, which allows for
non-universal, complex Wilson coefficients. 

As a second possibility, we examined new-physics contributions from
effective interactions of the form $(\bar{d}b)(\bar\tau\tau)$.  Depending on
their Lorentz structure such operators contribute at tree level to the decays $
B_d \to \tau^+ \tau^-$, $ B \to X_d \tau^+ \tau^-$ and $ B^+ \to \pi^+ \tau^+
\tau^+$. While for $B_d \to \tau^+ \tau^-$ there exists a direct experimental
bound, the branching ratios of $ B \to X_d \tau^+ \tau^-$ and $ B^+ \to \pi^+
\tau^+ \tau^-$ can only be bounded indirectly by using the lifetime difference
of $B_d$ and $B_s$ mesons. Loop-induced contributions to decays like $ B \to X_d
\gamma$, $ B^+ \to \pi^+ \mu^+ \mu^-$ and $B_d \to \gamma \gamma$ also arise
from $(\bar{d}b)(\bar\tau\tau)$ operators, but turn out to be relevant only for
tensor interactions.  Since the existing experimental constraints are all fairly
weak, we found that enhancements of both $\Delta \Gamma_d$ and $a_{sl}^d$ by
more than 100\% are possible also in the case of non-standard $b \to d \tau^+
\tau^-$ effects, in particular, if these new contributions stem from vector
operators. Given that some of the studied $B_{d,s}$-meson decays are expected to
be better measured or bounded in the near future by LHCb and eventually also
Belle II, we also presented
plots that illustrate the sensitivity of the individual decay channels to the
different Wilson coefficients. These results should prove useful in
monitoring the impact that further improved determinations of $ B_d \to \tau^+
\tau^-$, $ B \to X_d \tau^+ \tau^-$ and $ B^+ \to \pi^+ \tau^+ \tau^+$ have in
extracting information on $\Delta \Gamma_d$ .

We believe that the decay rate difference $\Delta \Gamma_d$ provides a unique
way to probe ``exotic'' new physics that leads to composite dimension-six
operators like $(\bar d c) (\bar c b)$ or $(\bar d b) (\bar \tau \tau)$. While
explicit new-physics realisations that give rise to such interactions are
difficult to construct, based on existing observations the possibility that
non-standard effects of this type enhance~$\Delta \Gamma_d$ and in this way
explain the observed anomalously large dimuon asymmetry $A_{\rm CP}$ cannot be
excluded. Direct measurements of $\Delta \Gamma_d$, more precise data on the
individual semi-leptonic CP asymmetries $a_{sl}^{d,s}$ and the lifetime
difference $\tau_{B_s} / \tau_{B_d}$, improvements in the experimental
determinations of rare and radiative $b \to d$ decays, but also a better
knowledge of the values of the CKM matrix elements $V_{cd}$ and $V_{cs}$ would
shed light on this issue. Such measurements (see also \cite{Gershon:2010wx})
should hence be pursued with vigour.

\section*{Acknowledgements}

We would like to thank Guido Bell, Guennadi Borissov, Tim Gershon and Roman Zwicky for helpful
discussions.  CB has been financed in the context of the ERC Advanced Grant
project ``FLAVOUR''(267104). UH acknowledges the warm hospitality and support of
the CERN theory division. GTX gratefully acknowledges the financial support of CONACyT 
(Mexico).

%
%

\appendix

\section{SM result for $\bm{\Delta\Gamma_d}$}
\label{app:DG}

Within the framework of the HQE, the off-diagonal element $\Gamma_{12}^d$ of the
decay rate matrix can be expressed as a double expansion in the inverse of the
heavy bottom-quark mass and in the strong coupling constant
\beq 
  \Gamma_{12}^d = \sum_{i=3}^\infty \sum_{j=0}^\infty 
  \left (\frac{\Lambda_{\rm QCD}}{m_b} \right )^i \left ( \frac{\alpha_s}{4 \pi} \right)^j \Gamma_i^{d, (j)} \,. 
\eeq
The terms $\Gamma_3^{d, (j)}$ encode the contributions from dimension-six
operators, whose matrix elements are parameterised in terms of the decay
constant $f_{B_d}$ and the bag parameters. The leading perturbative part
$\Gamma_{3}^{d,(0)}$ is already known since the 80's, while the NLO corrections
$\Gamma_{3}^{d,(1)} $ were calculated in
\cite{Beneke:1998sy,Beneke:2003az,Ciuchini:2003ww}. In the sub-leading HQE corrections
$\Gamma_4^{d,(j)}$, dimension-seven operators appear. Some of them can be
rewritten in terms of dimension-six operators, but others have to be estimated
in the vacuum insertion approximation (VIA). A first step in the
non-perturbative determination of the dimension-seven operators has been made
in~\cite{Mannel:2007am,Mannel:2011zza}. The perturbative contributions
$\Gamma_{4}^{d,(0)}$ were calculated in \cite{Dighe:2001gc}. Even the
sub-sub-leading corrections $\Gamma_{5}^{d,(0)} $ were estimated in
\cite{Badin:2007bv}. Because of our ignorance concerning the size of the matrix
elements of some of the appearing dimension-eight operators we will not use the
latter estimates, which in any case result in small corrections only.

\begin{center}
\begin{table}[ht]
\renewcommand{\arraystretch}{1.1}
\begin{tabular}{|lccc|lccc|}
\hline
  Parameter
& Value
& Unit
& Ref.
& Parameter
& Value
& Unit
& Ref.
\\
\hline \hline
\multicolumn{8}{|l|}{\textbf{Masses and couplings}}
\\
\hline
  $m_\mu$         & $0.105$                & GeV        & \cite{PDG:2013zta}
& $G_F$           & $1.16638\cdot 10^{-5}$ & GeV$^{-2}$ & \cite{PDG:2013zta}
\\ 
  $m_\tau$        & $1.777$                & GeV        & \cite{PDG:2013zta}
& $\alpha_s(M_Z)$ & $0.1184$               &            & \cite{PDG:2013zta}
\\
  $M_Z$           & $91.1876$              & GeV        & \cite{PDG:2013zta}
& $\alpha (M_Z)$ & $1/127.944$            &            & \cite{PDG:2013zta}
\\
  $M_W$           & $80.385$                 & GeV        & \cite{PDG:2013zta}
& $\alpha (m_b)$ & $1/132$                &            &
\\
\hline\hline
\multicolumn{8}{|l|}{\textbf{CKM}}
\\
\hline
  $\lambda$  & $0.22457^{+0.00186}_{-0.00014}$ &            & \cite{Charles:2004jd}
& $\gamma$   & $70 \pm 10$                     & ${}^\circ$ & \cite{Charles:2004jd}
\\ 
  $\rho$     & $0.1289^{+0.0176}_{-0.0094}$    &            & \cite{Charles:2004jd}
& $\eta$     & $0.348\pm 0.012$                &            & \cite{Charles:2004jd}
\\
  $|V_{td}^\ast V_{tb}^{} |$ & $0.00874$           &            &
&
  $|V_{cb}^{}|$ & $0.0416$  &            &
\\
  $|V_{ud}|$ & $0.9745$                        &            & \cite{Charles:2004jd}
& & & & 
\\
\hline\hline
\multicolumn{8}{|l|}{\textbf{Quark masses}}
\\
\hline
  $m_d$            & $4.8$              & MeV &
& $\m_c (\m_c)$             & $1.275\pm 0.025$ & GeV & \cite{PDG:2013zta}
\\
  $m_b^{\rm pole}$ & $4.8$            & GeV &
& $\m_b(\m_b)$       & $4.2$            & GeV &
\\ 
  $m_t^{\rm pole}$ & $173.1\pm 0.9$   & GeV & \cite{PDG:2013zta}
& $\m_t(\m_t)$       & $163.5$          & GeV &
\\
\hline\hline
\multicolumn{8}{|l|}{\textbf{$\bm{B}$-meson and light meson properties }}
\\
\hline
  $M_{B_u}$           & $5279.26$       & MeV & \cite{PDG:2013zta}
& $\tau_{B_u}$        & $1.641$         & ps  & \cite{PDG:2013zta}
\\
  $M_{B_d}$           & $5279.58$       & MeV & \cite{PDG:2013zta}
& $\tau_{B_d}$        & $1.519$         & ps  & \cite{PDG:2013zta}
\\
  $M_{B_s}$           & $5366.77$       & MeV & \cite{PDG:2013zta}
& $\tau_{B_s}$        & $1.463$         & ps  & \cite{PDG:2013zta}
\\
  $f_{B_{u,d}}$       & $190.5 \pm 4.2$ & MeV & \cite{Aoki:2013ldr}
& $f_{B_s}$           & $227.7 \pm 4.5$ & MeV & \cite{Aoki:2013ldr}
\\
  $f_{\pi}$           & $130.4$         & MeV & \cite{PDG:2013zta}
& $\Lambda_B$         & $400$           & MeV & \cite{Bell:2009fm}
\\
  $B_V$             & $0.84 \pm 0.07$ & GeV & \cite{Aoki:2013ldr}
& $B_S$             & $1.36 \pm 0.14$ & GeV & \cite{Yamada:2001xp}
\\
  $\widetilde{B}_S$ & $1.44 \pm 0.16$ & GeV & \cite{Becirevic:2001xt}
& & & & 
\\
\hline
\end{tabular}
\renewcommand{\arraystretch}{1.0}
  \caption{Collection of our input parameters for the numerical evaluation.}
  \label{tab:numeric:input}
\end{table}
\end{center}

Including the corrections mentioned above, we find the following SM prediction
for the width difference in the $B_d$-meson system
\bea
\begin{split}
\left ( \Delta \Gamma_d \right )_{\rm SM} & =  0.0029 \; \mbox{ps}^{-1} \bigg ( 1 
                    \pm 0.16_{B_{R_2}}
                    \pm 0.14_{f_{B_d}}
                    \pm 0.07_{\gamma}
                    \pm 0.07_{\mu} 
                    \pm 0.05_{\tilde{B}_S}
                    \pm 0.04_{B_{R_0}} \hspace{6mm}
 \\
               & \hspace{2.65cm}
                    \pm 0.03_{V_{cb}}
                    \pm 0.03_{B_V}
                    \pm 0.01_{m_b}
                    \pm 0.01_{m_c/m_b}
                    \pm 0.01_{|V_{ub}/V_{cb}|}
\bigg ) \\[2mm]
& =  (0.0029 \pm 0.0007) \; \mbox{ps}^{-1} \,.
\label{DGdSM}
\end{split}
\eea
In order to obtain this result we have used the same numerical input as in
\cite{Lenz:2011ti}. We see that combining the individual sources of uncertainty
in quadrature, $ \Delta \Gamma_d$ is known with a precision of around $25\%$
within the SM.

The dominant uncertainty in (\ref{DGdSM}) arises from matrix elements of
dimension-seven operators denoted by $R_2$ and $R_0$ (see \cite{Lenz:2006hd} for
the precise definitions). These contributions gives rise to an uncertainty of
$16\%$ and $4 \%$, respectively.  One should keep in mind that the corresponding
bag parameters have been estimated in \cite{Lenz:2011ti} quite conservatively by
allowing a $50\%$ deviation from the VIA value of 1. Allowing only for
modifications of $25\%$, would reduce the errors corresponding to
dimension-seven operators by a factor of 2.  While the bag parameters of some of
the dimension-seven operators have been estimated in
\cite{Mannel:2007am,Mannel:2011zza} using QCD sum rules, a lattice calculation
of $B_{R_2}$ and $B_{R_0}$ is unfortunately not available at present.  Any
progress in this direction would be of utmost importance to reduce the total
uncertainty in $\left ( \Delta \Gamma_d \right )_{\rm SM} $.

The second largest uncertainty in (\ref{DGdSM}) arises from the matrix elements
of dimension-six operators. An error of $14 \%$ comes from the decay constant
$f_{B_d}$ and an uncertainty of $5 \%$ ($3\%$) stems from the bag parameter
$\tilde{B}_S$ ($B_V$).  Several lattice groups are working on the determination of
these hadronic parameters, so that these errors are expected to shrink.

Number three in the error budget of $\left ( \Delta \Gamma_d \right )_{\rm SM} $
are the CKM uncertainties. The error due to $\gamma$ ($V_{cb}$) amounts to $7
\%$ ($3 \%$) at present, but is expected to improve in the future. The
renormalisation scale ($\mu$) dependence gives rise to an uncertainty of about
$7 \%$. To reduce this error a NNLO order calculation would be necessary. The
remaining parametric uncertainties in (\ref{DGdSM}) are negligible at the
current stage.

%
%
\section{Numerical input}
\label{app:numeric:input}

In Table \ref{tab:numeric:input}
 we collect the numerical values of various parameters used as
input throughout the work. The decay constants $f_{B_q}$ are the most recent
averages of the FLAG compilation \cite{Aoki:2013ldr} that incorporate the $N_f =
2 + 1$ results of \cite{Bazavov:2011aa, McNeile:2011ng, Na:2012kp}. More recent
determinations with $N_f = 2 + 1 + 1$~\cite{Dowdall:2013tga} and $N_f =
2$~\cite{Carrasco:2013zta} are consistent with these averages.

%
%

%
%



\end{document}